\documentclass[review, 3p]{elsarticle}

\usepackage{subcaption}
\usepackage{amsmath}
\usepackage{amssymb}
\usepackage{graphicx}
\usepackage{color}
\usepackage{morefloats}
\usepackage{multirow}
\usepackage{hyperref}
\usepackage{algorithm}
\usepackage{algorithmic}

\newcommand{\Rmnum}[1]{\MakeUppercase{\romannumeral #1}}

\begin{document}
\begin{frontmatter}

\title{ADWISERv2: A Plug-and-play Controller for Managing TCP Transfers in IEEE~802.11 Infrastructure WLANs with Multiple Access Points}
\tnotetext[t1]{The research reported in this paper was conducted under a project funded
by the Department of Electronics and Information Technology, Government
of India, and was reported in part in a demonstration abstract at
Department of Electrical Communication Engineering, Indian Institute of
Science, Bangalore, India}

\author[dese]{Albert Sunny \corref{mycorrespondingauthor}}
\cortext[mycorrespondingauthor]{Corresponding author}
\ead{salbert@dese.iisc.ernet.in}

\author[ece]{Sumankumar Panchal}
\ead{suman@ece.iisc.ernet.in}


\author[ece]{Nikhil Vidhani}
\ead{nvidhani@cisco.com}

\author[ece]{Subhashini Krishnasamy}
\ead{subhashini.kb@gmail.com}

\author[ece]{S.V.R. Anand}
\ead{anand@ece.iisc.ernet.in}

\author[ece]{Malati Hegde}
\ead{malati@ece.iisc.ernet.in}

\author[dese]{Joy Kuri}
\ead{kuri@dese.iisc.ernet.in}

\author[ece]{Anurag Kumar}
\ead{anurag@ece.iisc.ernet.in}

\address[dese]{Department of Electronic Systems Engineering, Indian Institute of Science, Bangalore, India}
\address[ece]{Department of Electrical Communication Engineering, Indian Institute of Science, Bangalore, India}

\begin{abstract}
   In this paper, we present a \emph{generic plug-and-play controller} that ensures fair and efficient operation of IEEE~802.11 infrastructure  wireless local area networks with multiple co-channel access points, without any change to hardware/firmware of the network devices.  Our controller addresses  performance issues of TCP transfers in multi-AP WLANs, 
  by overlaying a coarse time-slicing scheduler on top of a cascaded fair queuing scheduler. 
  The time slices and queue weights, used in our controller, are obtained from the
  solution of a constrained utility optimization formulation. A study of the impact of coarse time-slicing on TCP is also presented in this paper. We present an improved algorithm for adaptation of the service rate of the fair queuing scheduler and provide experimental results to illustrate its efficacy. We also present the changes that need to be incorporated to the proposed approach, to handle short-lived and interactive TCP flows. Finally, we report the results of experiments performed on a real testbed, demonstrating the efficacy of our controller.
\end{abstract}

\begin{keyword}
QoS Management; Centralized Fair Queuing Scheduler; 802.11 WLANs; Coarse Time-Slicing
\end{keyword}

\end{frontmatter}


\section{Introduction and Related Work}
\label{sec:introduction}

The widespread use of  IEEE~802.11 infrastructure Wireless Local Area
Networks (WLANs) has enabled mobile users to
seamlessly transfer huge volumes of data. While IEEE~802.11 infrastructure WLANs provide mobility, and are a cheap alternative to cellular networks, they
are well known to display several performance anomalies \cite{survey.ahmed-keshav06smarta,survey.shrivastava-etal09centaur,adwiser_ton}, for example, multi-rate
unfairness, uplink-downlink unfairness, hidden and exposed node problems.
Thus, there is a need to study such performance anomalies, and provide better performance
management solutions for these infrastructure WLANs.

In \cite{survey.magistretti-etal10mitigating-link-hindering-transmissions-80211},
Magistretti et al. introduce MIDAS, a management framework that addresses under-served
links by throttling traffic to the interfering links. Based on the notion of ``Activity Share'' that they
introduce, the authors propose a method to assess the potential throughput gain
that the under-served link can experience, when hindering transmissions are
rate-limited. MIDAS limits the rate at which traffic
is sent to the interfering links, so that the under-served link's
activity share improves. While MIDAS improves the ``air time'' of under-served links, there are no flow-level system objectives that drive the target activity share. Such flow-level system objectives are indispensable in networks with Quality-of-Service (QoS) guarantees.

SMARTA \cite{survey.ahmed-keshav06smarta} is a centralized controller that sets and dynamically adjusts the operating parameters of enterprise WLAN Access Points (e.g., channel frequency and power allocation) to optimize a predefined objective. In \cite{survey.ahmed-keshav06smarta}, the authors describe several simple tests to estimate the interference environment and obtain a \emph{Conflict Graph}.   Based on this conflict-graph, the controller executes channel assignment and power control algorithms that aim to minimize the number of conflicting transmissions. We note that SMARTA focuses on system configuration,  and that there are no mechanisms for controlling the rate of traffic on the links. Also, SMARTA does not incorporate the capacity of the wide-area-network (WAN)-WLAN links into its design. Therefore, SMARTA may not perform efficiently in networks with WAN-WLAN traffic.  Further, we would like to note that SMARTA is evaluated via simulations and not on a real testbed. 

CENTAUR \cite{survey.shrivastava-etal09centaur} is motivated by the observation
that when the traffic pattern is dominated by downloads, the IEEE~802.11 distributed coordination function (DCF) leads to wasted airtime in networks with hidden and exposed terminals. To tackle this, the authors in
\cite{survey.shrivastava-etal09centaur} propose a centralized controller
for hidden and exposed terminal links; the objective is to ensure that transmissions to hidden nodes do not happen simultaneously,
while transmissions to exposed nodes overlap, in time,  as much as possible. Links that are not associated with hidden or exposed nodes use the IEEE~802.11 DCF to access the medium. Handling exposed terminals requires carrier sensing to be effectively disabled. To achieve this,  the authors in \cite{survey.shrivastava-etal09centaur} use fixed backoffs at the access points (APs)  --- this requires modifications to the AP firmware.  Since most firmwares are not open-source, it is difficult to incorporate such modifications.

\begin{figure}[ht]
	\centering
	\includegraphics[scale=0.25]{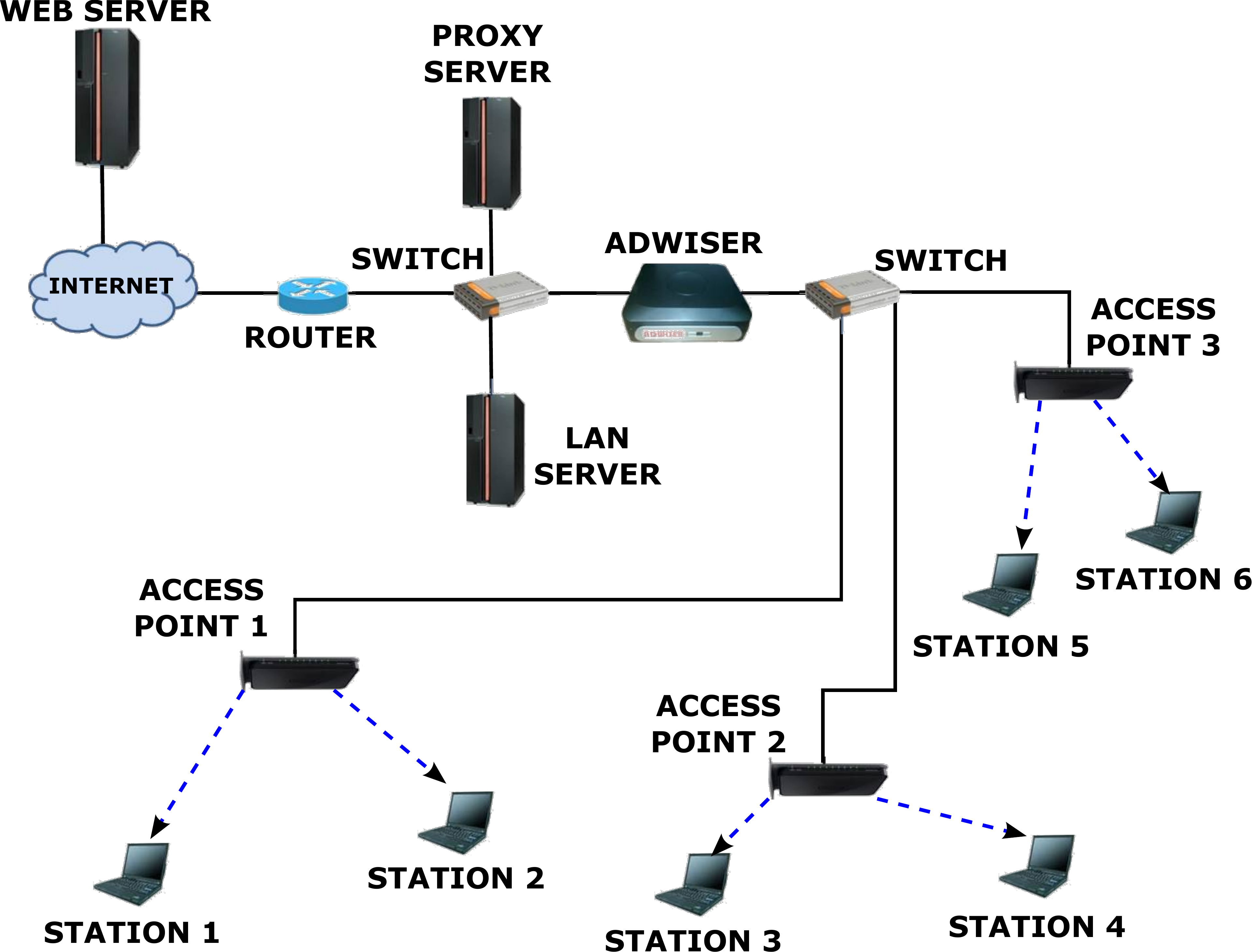}
	\caption{A multi-AP enterprise network managed by ADWISER. \label{fig:multi-ap-testbed}}
\end{figure}

In our earlier work \cite{adwiser_ton}, we had introduced ADWISER (\textbf{AD}vanced \textbf{W}iFi
\textbf{I}nternet \textbf{S}ervice \textbf{E}nhance\textbf{R}); a \emph{plug-and-play centralized} device for WLAN performance management. ADWISER is located between the WLAN and LAN (see Fig. \ref{fig:multi-ap-testbed}) so that all packets to and from the wireless STAs pass through it.  ADWISER uses virtual servers and queues to pull into itself the queues from two bottleneck resources: the WLAN medium and the Internet access link. As a result, ADWISER was able to mitigate infrastructure WLAN performance anomalies, such as throughput inefficiency due to multi-rate association, unfairness between downlink and uplink TCP controlled file transfers, and throughput unfairness between intranet and Internet downloads. \emph{However,  ADWISER was not equipped to manage networks with interfering co-channel APs.}

\section{Our Contributions} \label{sec:our_contri}

In this paper, in order to address the performance issues for TCP transfers in a multi-AP setting, we propose ADWISER~v2, a \emph{generic plug-and-play} controller which overlays \emph{coarse time-slicing} over the queuing and scheduling architecture of ADWISER. While the concerns that have driven our research into the second version of ADWISER are similar to those expressed in \cite{survey.magistretti-etal10mitigating-link-hindering-transmissions-80211, survey.ahmed-keshav06smarta, survey.shrivastava-etal09centaur}, our main goal was to provide a \emph{generic plug-and-play} controller; a controller requiring no changes to the hardware or firmware of APs and clients. The key idea behind ADWISER~v2 is to divide time into large slices, and schedule packets to non-interfering links in the same slice.  While the ``epochs'' in CENTAUR and the ``time slices" in ADWISER~v2 are similar, the biggest difference is that ADWISER~v2 can be used without modifications to the firmware/hardware of the APs and clients. Table~\ref{table:summary} compares a few  features of the WLAN controllers discussed in this paper.  

\begin{table}[h]
	\centering
	\caption{Table comparing the WLAN controllers discussed in this paper. \label{table:summary}}
	\begin{tabular}{|c|c|c|c|c|c|}
		\hline
		\multirow{2}{*}{Features}  & \multirow{2}{*}{CENTAUR}  & \multirow{2}{*}{SMARTA} & \multirow{2}{*}{MIDAS} & \multirow{2}{*}{ADWISER} & \multirow{2}{*}{ADWISER~v2} \\ 
		& \cite{survey.shrivastava-etal09centaur} & \cite{survey.ahmed-keshav06smarta} & \cite{survey.magistretti-etal10mitigating-link-hindering-transmissions-80211} &\cite{adwiser_ton} &\\
		\hline well defined QoS objectives   & $\times$  & $\times$  & $\times$  & \checkmark & \checkmark \\ 
		\hline no modification to APs    & $\times$ &  $\times$ & $\times$ & \checkmark & \checkmark \\ 
		\hline no modification to clients     & \checkmark & \checkmark  &  \checkmark  &  \checkmark & \checkmark \\ 
		\hline evaluated on a real test-bed    &  \checkmark &  $\times$ & \checkmark  & \checkmark & \checkmark \\ 
		\hline consider interfering co-channel APs    & \checkmark & \checkmark  & \checkmark  & $\times$  & \checkmark \\ 
		\hline 
	\end{tabular} 
\end{table}

The main contributions of this paper are as follows:

\begin{itemize}
\item A study, design, and demonstration of coarse-time slicing, for managing long-lived TCP controlled transfers in WLANs with interfering co-channel APs.
\item Detailed experimental study of the performance of TCP with time-slicing, thereby providing essential support for the idea of coarse grained time-sliced
scheduling of the WLAN medium. 
\item Formulation of the problem of rate allocation to the clients, subject to the constraints of enterprise networks.
\item An approach for inferring link dependencies in the network, and a rate adaptation algorithm for long-lived file transfers.
\item Experimental results that demonstrate the effectiveness of our approach in achieving the desired  resource
sharing objectives.
\item Discussions on short-lived and interactive TCP traffic, and a methodology for their management.
\end{itemize}

The remainder of the paper is organized as follows. In Section~\ref{sec:motivation}, we discuss the motivation for
coarse time-sliced scheduling. The effect of coarse timeslicing on TCP transfers is
discussed in Section~\ref{sec:effect_of_timeslicing}.  In Section \ref{sec:utility-optimization}, we
formulate a constrained utility optimization problem from which the time slices
are derived. An online, low overhead, heuristic 
that infers the dependence graph is presented in Section \ref{sec:dynamic_link_dependency}. Section \ref{sec:rate_adaptation}
presents an algorithm to adaptively estimate the sustainable service rate of
long-lived TCP transfers.  Experiments that demonstrate the usefulness of our approach are
presented in Section \ref{sec:experimental_results}. In Sections \ref{sec:sl}
and \ref{sec:it}, we discuss the management of short-lived and interactive TCP
traffic, respectively. An experiment with IEEE~802.11n infrastructure WLAN is presented 
in Section~\ref{sec:11n}. Finally, in Section \ref{sec:conclusion}, we conclude
the paper.

\section{Motivation}
\label{sec:motivation}

Packet-by-packet scheduling and fine-grained time-slicing require high resolution timers with stringent real-time scheduling from the operating system \cite{survey.shrivastava-etal09centaur}. During heavy network traffic, such scheduling could result in starvation of the controller resources. Also, in multi-AP WLANs, fine-grained time-slicing (typically tens of milliseconds) requires tight coordination between the centralized controller and the access points, to ensure conflict free transmission \cite{survey.shrivastava-etal09centaur}. Meeting such stringent requirements would inevitably require the \emph{employment of customized access points.} On the other hand, coarse time-slicing can be implemented without any modification to the clients or access points. In coarse time-slicing, time is divided into ``time-frames.'' We use the adjective ``coarse'' to describe the time slices, because many packets flow during  each  slot, there are  no packet by packet controls. 
A time-frame is further subdivided into slots, and only non-interfering links are allowed to transmit in any given slot, thereby mitigating interference. Due to its low scheduling overhead, coarse time-slicing can support a large number of
APs and clients on a single machine, even during periods of heavy network traffic. All the above reasons make our approach
attractive in terms of implementability, cost effectiveness, and scalability.

\begin{figure}[ht]
\centering
\includegraphics[height=30mm]{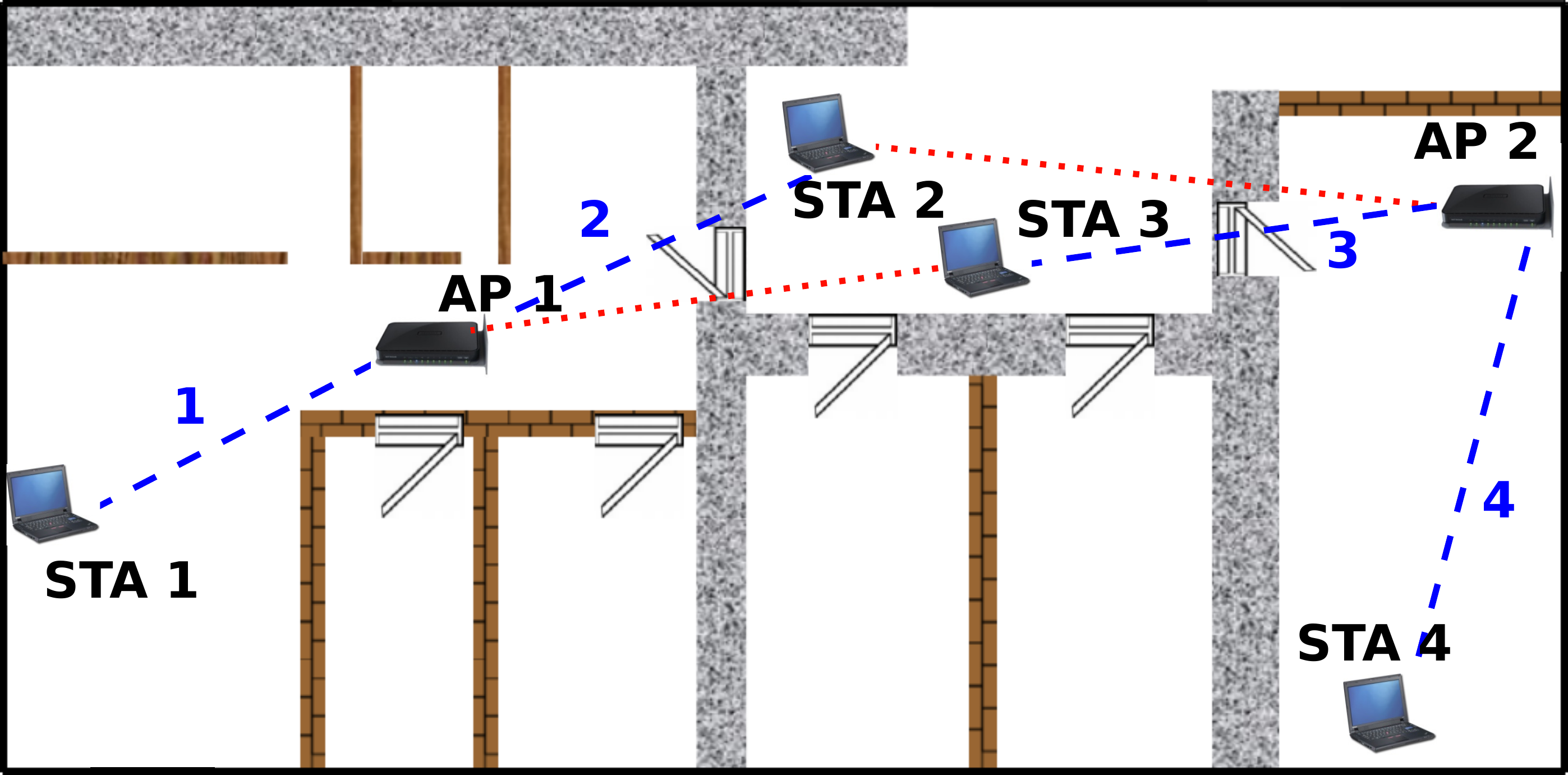}
\caption{A WLAN with 2 APs and 4 clients. The dashed lines indicate client-AP associations,
    while the dotted lines indicate client-AP interference. The legend for the building material is provided in Fig.~\ref{fig:ece_layout_for_paper}.
    \label{fig:multi_ap_four_sta_exp_layout} }
\end{figure}

We demonstrate the simplicity and effectiveness of coarse time-slicing by doing an experiment on the setup depicted in Fig.~\ref{fig:multi_ap_four_sta_exp_layout}. In this setup, there are four clients (STAs)  associated with two co-channel IEEE~802.11g APs at a physical rate of $54 \, Mbps$, and each client is downloading a large file from a server on the local area network. 

\begin{figure}[ht]
\centering
\includegraphics[scale=0.4]{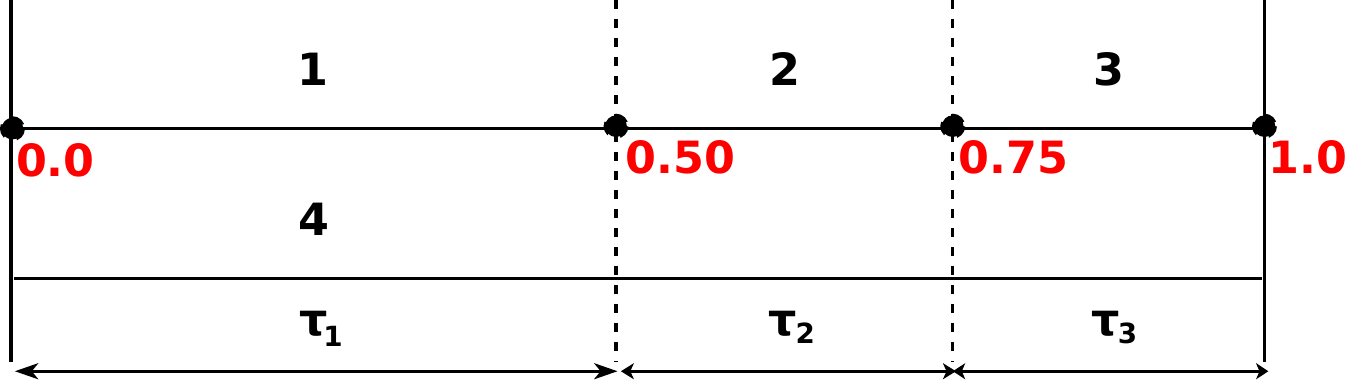}
\caption{2 APs and 4 clients; LAN transfers to the clients (experimental setup in Fig.~\ref{fig:multi_ap_four_sta_exp_layout}): Time slices allocated to the clients in a time-frame of duration $1000 \, ms$.}
\label{fig:four_sta_exp_1_time_slicing}
\end{figure}

\begin{figure}[ht]
  \centering
  \includegraphics[scale=0.35,angle=-90]{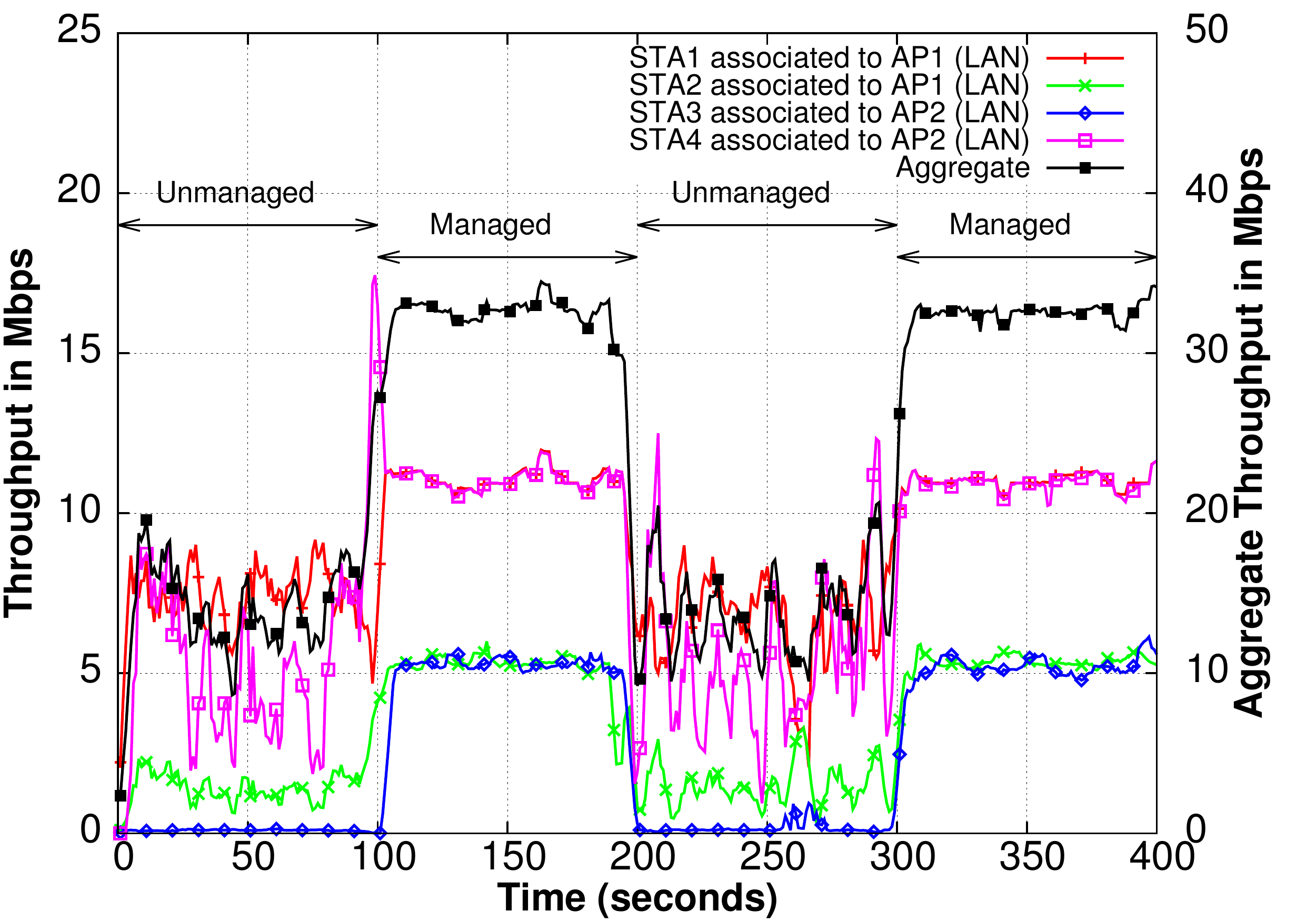}
  \caption{2 APs and 4 clients (STAs); LAN transfers to the clients (experimental setup in Fig.~\ref{fig:multi_ap_four_sta_exp_layout}): Individual and aggregate throughputs of the clients.}
  \label{plot:multi_ap_four_sta_all_lan_exp}
\end{figure}

We divide time into time-frames of duration $1000 \, ms$. Within each frame, we
allocate time slots to the clients in the network (see
Fig.~\ref{fig:four_sta_exp_1_time_slicing}). In the first slot, STA~1 and STA~4
are scheduled together for a duration of $500 \, ms$ each. In the next slot,
STA~2 is scheduled for a duration of $250 \, ms$. Finally, in the last slot,
STA~3 is scheduled for a duration of $250 \, ms$.   To compare the performance of
coarse time-slicing and the default IEEE~802.11 DCF behavior, we operate
alternately in two modes: ``unmanaged mode''  where we disable timeslicing, and
``managed mode'' where timeslicing is enabled.
Fig.~\ref{plot:multi_ap_four_sta_all_lan_exp} shows the TCP throughputs obtained by
the clients (scale is on the left side of Fig.~\ref{plot:multi_ap_four_sta_all_lan_exp}) and the aggregate TCP throughput (scale is on the right side of Fig.~\ref{plot:multi_ap_four_sta_all_lan_exp}), in each of the modes.

During the intervals $0-100 \, seconds$ and $200-300 \, seconds$, the network
is in \emph{unmanaged mode} and the throughputs obtained indicate the behaviour of the
default IEEE~802.11~DCF. With the four clients contending simultaneously, STA~3
obtains a very low throughput (almost zero). STA~1, STA~2 and STA~4 obtain highly
variable throughputs of about $7 \, Mbps$, $2 \, Mbps$ and $5 \, Mbps$,
respectively. STA~2 and STA~3  get very low throughput
because these are the links ``in-the-middle'' (hidden nodes). Also, STA~1 and
STA~4 obtain highly variable throughput, even though they do not interfere with each other. \emph{This experiment shows that in the unmanaged mode, the aggregate throughput of the network is unpredictable and highly variable.}

Experiments indicate that a client associated at physical rate of $54\, Mbps$ with an AP, downloading a large file obtains a maximum TCP
throughput of $23 \, Mbps$ \cite{adwiser_ton}. Therefore, in the \emph{managed mode}, in this experiment, ADWISER~v2 serves packets at a fixed rate of $22 \,Mbps$ (refer Section~\ref{sec:rate_adaptation} for a discussion on dynamically setting ADWISER~v2 service rate). As
can be seen from the Fig.~\ref{plot:multi_ap_four_sta_all_lan_exp}, during
the periods $100-200 \, seconds$ and $300-400 \,seconds$, STA~1 and STA~4 obtain
throughputs of $22 \cdot \frac{500}{1000} = 11 \, Mbps$ each, and STA~2 and STA~3
obtain $22 \cdot \frac{250}{1000} = 5.5 \, Mbps$, as expected. In addition, we see that, due to the
independent link scheduling, the throughputs are quite flat over time. Further,
the aggregate throughput increases to about $32 \, Mbps$. \emph{This is a remarkable improvement over
the unmanaged situation, all being achieved with no changes to the firmware/hardware of the APs or the
clients. }


\section{Effect of Time Slicing on TCP} \label{sec:effect_of_timeslicing}

\begin{figure}[ht]
	\centering
	\includegraphics[scale=0.45]{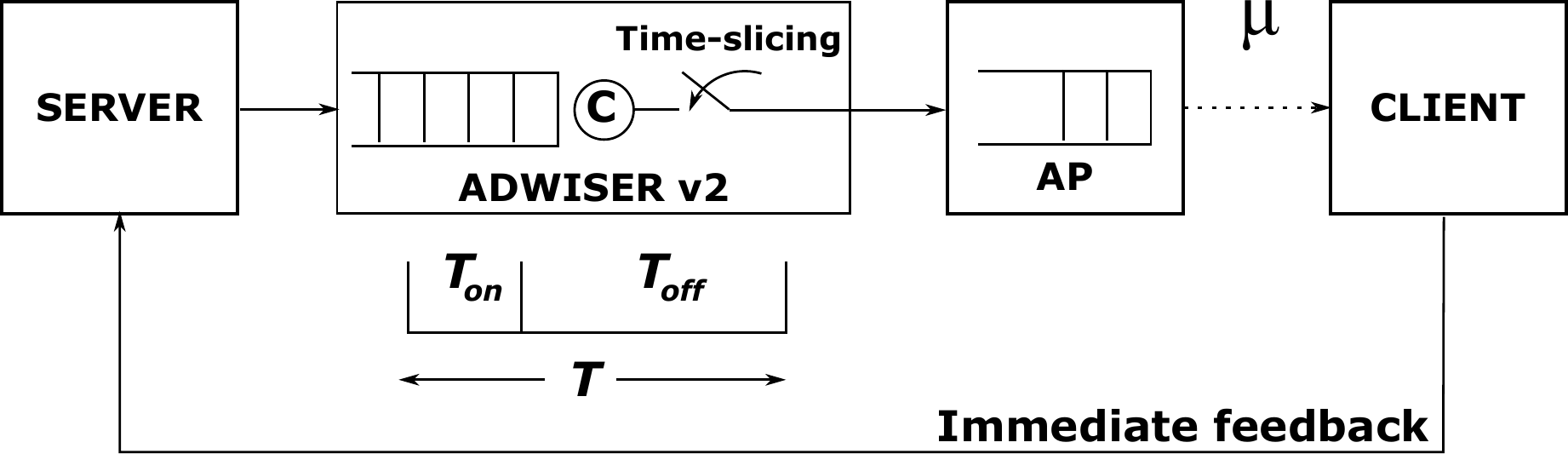}
	\caption{An abstracted model of time-sliced TCP connection between a LAN/WAN server and a WLAN client.}
	\label{fig:link_ts_wireless}
\end{figure}

An abstracted model of a time-sliced TCP connection between a server and a wireless client is shown in Fig.~\ref{fig:link_ts_wireless}. In Fig.~\ref{fig:link_ts_wireless}, the AP is depicted as a queue with a TCP service rate $\mu \, Mbps$. Due to random access, the service at the AP is stochastic, and $\mu$ denotes the average TCP service rate. ADWISER~v2 is depicted as a queue with a constant service rate of $C \, Mbps$. Over successive intervals of duration $T$, ADWISER~v2 serves the client for $T_{on}$ units of time at a rate of $C \, Mbps$ and abstains from service for $T_{off} = T - T_{on}$ units of time.  Since time-sliced TCP connections encounter links that seem to follow an \emph{ON-OFF} pattern; this could trigger timeouts, leading to poor 
throughputs in long-lived TCP transfers, and poor response times in short-lived (web-like transfer) and interactive traffic.

In this section, in detail, we study the effect of time-slicing on long-lived TCP transfers. The effect on short-lived and interactive traffic will be discussed in Section~\ref{sec:slit}. We restrict our study to the case when the server is running either ``TCP Cubic'' or ``TCP Reno.'' This choice is motivated by the fact that about $60\%$ of web servers with valid traces use one of these variants as their TCP congestion control algorithm  \cite{survey.yang-etal11ca-identification}. The hardware specifications of the network devices used for experiments in this section are provided in Section~\ref{sec:testbed}. To study the TCP congestion window, in this section, we have used the standard \emph{tcpprobe linux module}. Further, the SACK option of TCP was enabled in all our experiments. Time-sliced TCP connection, of a client on WLAN, will either be a LAN-WLAN connection or a WAN-LAN-WLAN connection. LAN-WLAN connections have low delays (order of milliseconds), since the server is located on the local area network. On the other hand, WAN-LAN-WLAN TCP transfers have large end-to-end delay (of the order of hundreds of milliseconds). current Linux systems implement an algorithm called Forward
RTO Recovery that attempts to detect a spurious timeout \cite{frto_rfc}. Thus, we will examine each case separately with and without F-RTO.  

\subsection{Time-sliced LAN-WLAN TCP Transfers}
\label{sec:tcp_perf_with_timeslicing}

\begin{figure}[ht]
	\centering
	\includegraphics[width=0.45\linewidth]{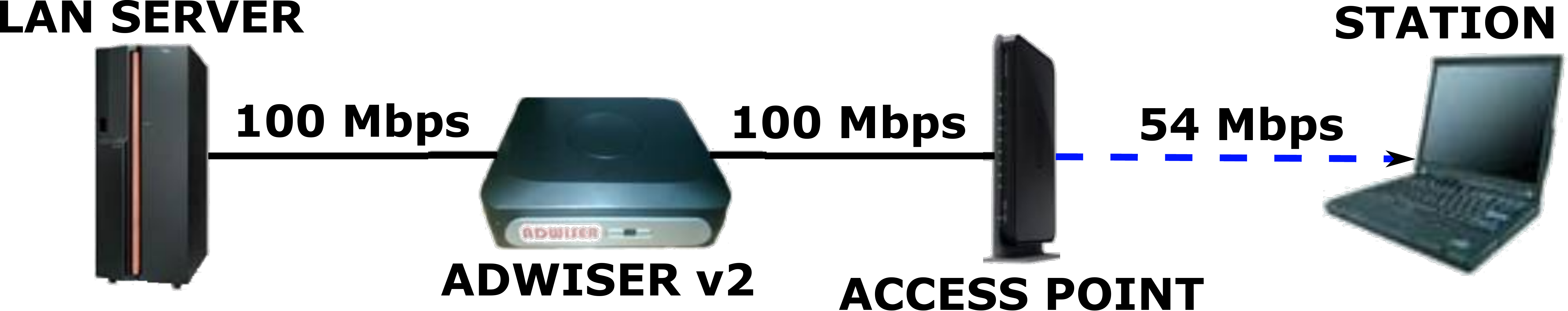}
	\caption{Experimental setup for LAN-WLAN transfers.}
	\label{fig:lan_exp_setup}
\end{figure}

In this section, we study the effect of time-slicing on LAN-WLAN TCP connections, i.e, for TCP controlled transfers between a server on the LAN and a client on on the WLAN. The experimental setup consists of a server, a wireless client, an IEEE 802.11g AP and ADWISER~v2 connected as shown in Fig~\ref{fig:lan_exp_setup}. We set the service rate $C$ of the virtual queue in ADWISER~v2 to $22 \, Mbps$. The client is associated with the AP at a physical rate of $54 \, Mbps$ (this corresponds to an average TCP throughput of $\mu = 23 \, Mbps$). The average round-trip propagation delay (RTPD) between the server and the client is $0.5 \, ms$. \emph{Since ADWISER~v2 serves the queue at a rate of $22 \, Mbps$ and $0 \, Mbps$ for $T_{on}$ and $(T - T_{on})$ units of time respectively, we expect the
time-sliced throughput to be $22 \times \frac{T_{on}}{T} \, Mbps$.} To study the effect of different time slices on LAN-WLAN TCP connections, we performed long file downloads for various on-times ($T_{on}$). Table \ref{tables:lan_table} presents the results of our experiment for TCP Cubic and TCP Reno, with and without F-RTO.

\begin{table}[h]
\centering
  	\caption{Expected and measured (averaged over $10$ runs) throughput of time-sliced LAN-WLAN TCP connections; the time-frame duration $T$ is $1000 \, ms$, $C=22 \, Mbps$ and client is associated at $54 \, Mbps$.}
\begin{tabular}{|c|c|c|c|c|c|}
\hline 
\multirow{3}{*}{$T_{on}$} &  \multicolumn{4}{ c |}{Measured throughput (Mbps)} & \multirow{1}{*}{Expected}\\
\cline{2-5}
 &  \multicolumn{2}{c|}{Cubic} & \multicolumn{2}{c|}{Reno} & \multirow{1}{*}{throughput} \\
\cline{2-5}
 & F-RTO & No F-RTO & F-RTO & No F-RTO  &  (Mbps)\\
\hline 
$50 \,ms$   &  $1.08 \, $  & $1.10 \, $  &  $1.10 \, $  & $1.06 \, $ & $1.10 \, $\\
\hline
$200 \,ms$   &  $4.37 \, $  & $4.39 \, $  &  $4.35 \, $  & $4.40 \, $ & $4.40 \, $\\
\hline
$400 \,ms$   &  $8.77 \, $  & $8.78 \, $  &  $8.76 \, $  & $8.79 \, $ & $8.80 \, $\\
\hline
$600 \,ms$   &  $13.13 \, $  & $13.19 \, $  &  $13.15 \, $  & $13.19 \, $ & $13.20 \, $\\
\hline
$800 \,ms$   &  $17.60 \, $  & $17.60 \, $  &  $17.59 \, $  & $17.58 \, $ & $17.60 \, $\\
\hline
$1000 \,ms$   &  $22.00 \, $  & $22.00 \, $  &  $22.00 \, $  & $22.00 \, $ & $22.00 \, $\\
\hline
\end{tabular}
\label{tables:lan_table}
\end{table}

By comparing the expected and measured throughputs presented in 
Table~\ref{tables:lan_table}, we can conclude that time-slicing does
not cause degradation in TCP throughput of time-sliced LAN-WLAN TCP connections. This is due to the fact that during the on-time,
the low RTPD on the LAN allows the TCP sender to recover
from RTOs by quickly ramping up its congestion
window.

\subsection{Time-sliced WAN-LAN-WLAN TCP Transfers and the Need for a TCP Proxy}

\begin{figure}[ht]
	\centering
	\includegraphics[width=0.6\linewidth]{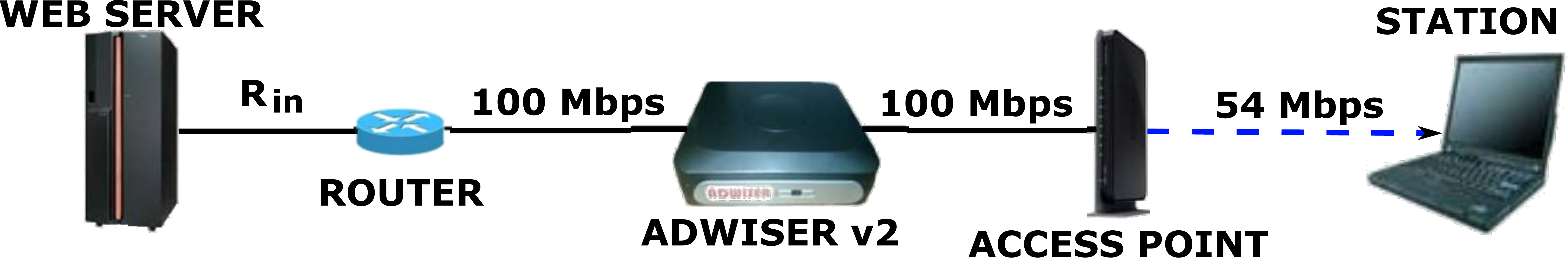}
	\caption{Experimental setup for WAN-LAN-WLAN transfers with RTPD of $150 \, ms$.}
	\label{fig:wan_expt_setup}
\end{figure}

In this section, we study the effect of time-slicing when the TCP
connection is transferring data between a server on the WAN and a client
on the WLAN. The experimental setup is shown in Fig.~\ref{fig:wan_expt_setup}. In this experiment, we emulate a WAN  with a round trip propagation delay of $150 \, ms$ and a WAN-LAN access link of capacity $R_{in} \, Mbps$. Our choice, of an RTPD of $150 \, ms$ for the WAN link, is motivated by a worst-case scenario. The RTPD is configured in the server by using Linux's \emph{netem} tool, and a commercial router has
been used for configuring the Internet access link speed. The client is associated with the access point at $54 \, Mbps$, and is downloading a large file from a server on the WAN. 

\begin{figure}[ht]
	\centering
	\includegraphics[scale=0.32]{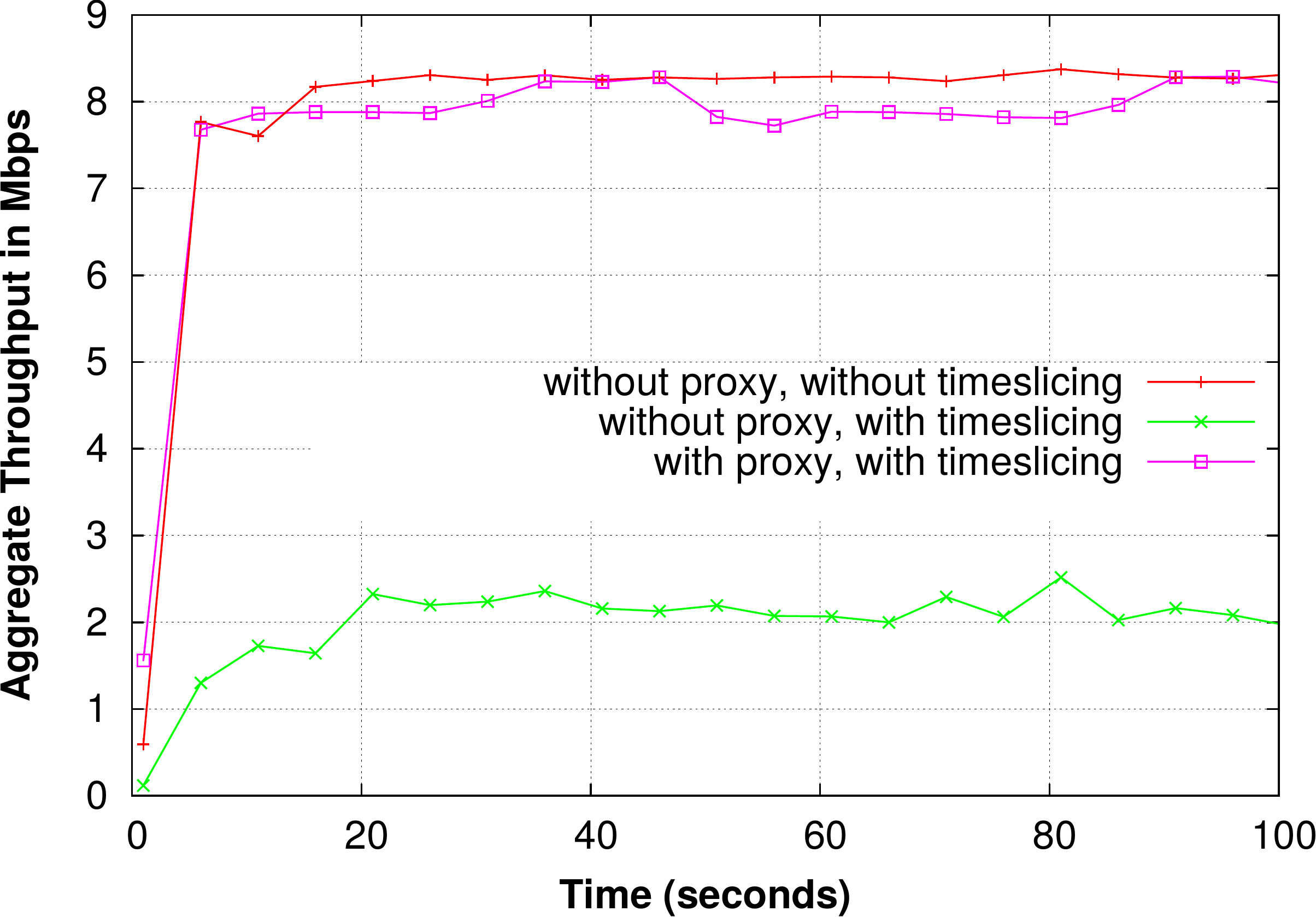}
	\caption{Throughput of a time-sliced WAN-LAN-WLAN TCP Cubic connection (experimental setup in Fig.~\ref{fig:wan_expt_setup}) when \textbf{WAN link is bottleneck}: $C = 22\, Mbps$, $T_{on} = 360 \, ms$, $T_{off} = 640 \, ms$, $\mathbf{R_{in} = 8 \, Mbps}$, and $RTPD = 150 \,ms$ \label{fig:throughput_cubic_wan_8}.}
\end{figure}

We first consider an $R_{in} = 8 \, Mbps$ WAN link and a client associated with the access point at a physical rate of $54 \, Mbps$. Measurements show that a client associated at a physical rate of $54 \, Mbps$ can sustain an average TCP throughput of about $23 \, Mbps$. Since the WAN link is the bottleneck in this scenario, we expect the client to obtain a throughput of $8 \, Mbps$ (see the plot for ``without proxy, without time-slicing'' in Fig.~\ref{fig:throughput_cubic_wan_8}). Now, if ADWISER~v2 is introduced with its virtual service rate $C$ set to $22 \, Mbps$, with a time-slicing of $T_{on} = 360 \,ms $ and $T_{off} = 640 \,ms $, we expect to obtain a TCP throughput of $22 \times \frac{T_{on}}{T_{on}+T_{off}} = 8 \, Mbps$. But, from Fig.~\ref{fig:throughput_cubic_wan_8}, we can see that without a proxy and with time-slicing, the client obtains a throughput of about $2 \, Mbps$, which is
roughly $25\%$ of the throughput achievable if the virtual server were busy all the time.

\begin{figure}[ht]
	\centering
	\includegraphics[scale=0.35]{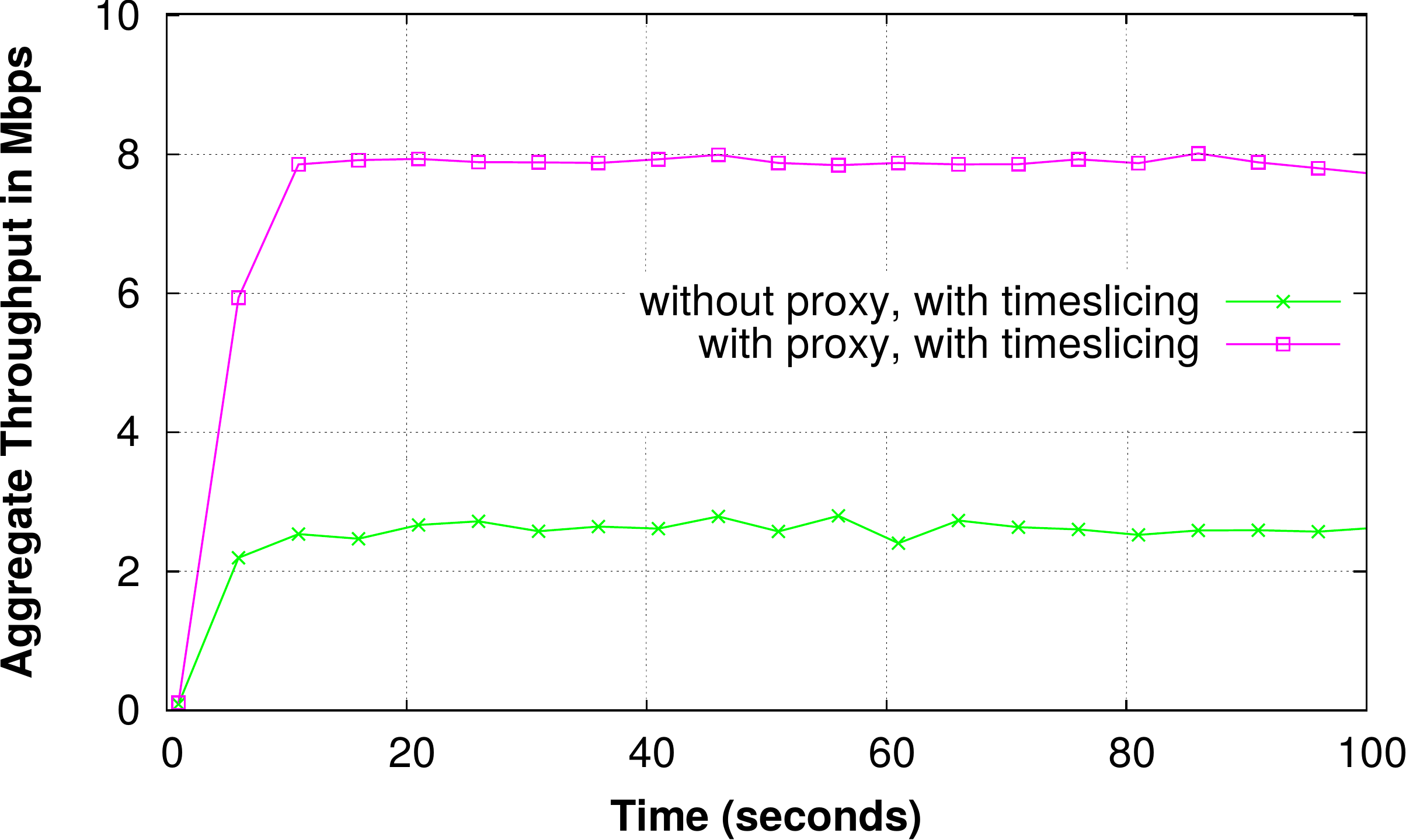}
	\caption{Throughput of a time-sliced WAN-LAN-WLAN TCP Cubic connection (experimental setup in Fig.~\ref{fig:lan_exp_setup}) when the \textbf{WLAN link is the bottleneck}: $C = 22\, Mbps$, $T_{on} = 360 \, ms$, $T_{off} = 640 \, ms$, $\mathbf{R_{in} = 32 \, Mbps}$, and $RTPD = 150 \,ms$. \label{fig:throughput_cubic_wan_32}}
\end{figure}

\begin{figure}[ht]
	\begin{subfigure}[b]{\textwidth}
		\centering
		\includegraphics[scale=0.28]{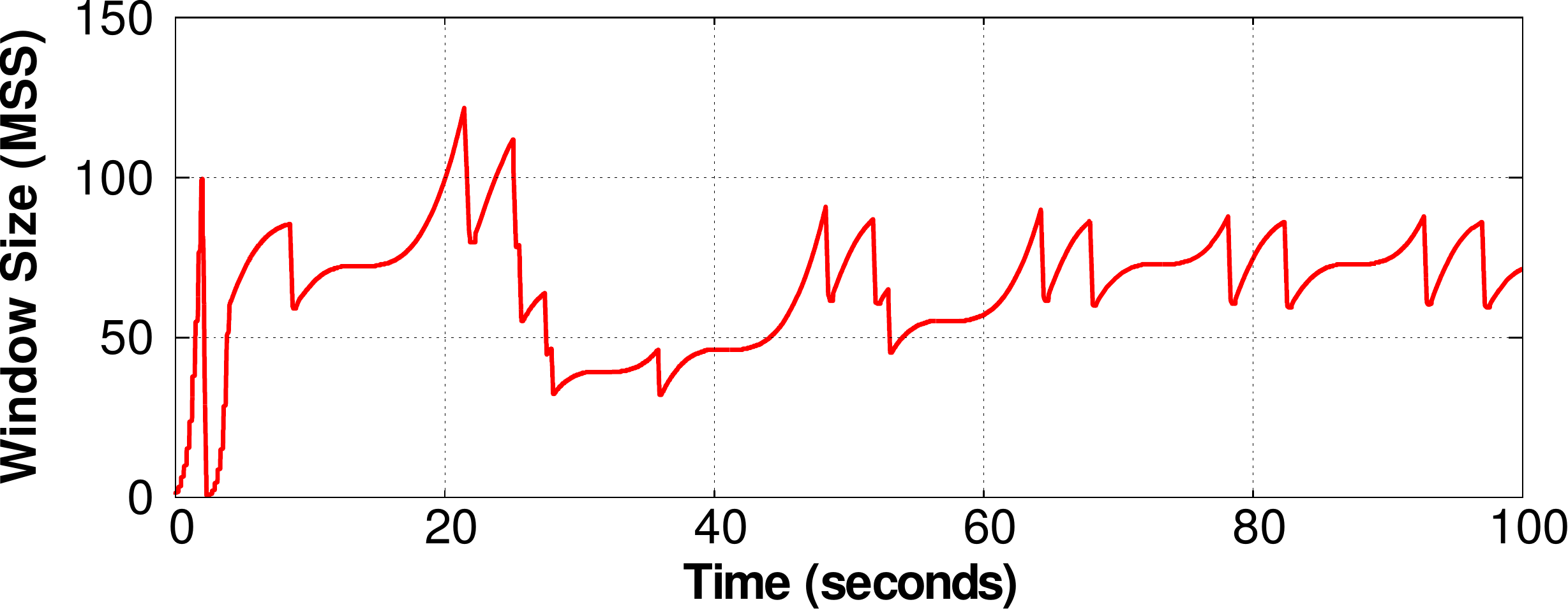}
		\caption{Without proxy, without time-slicing {\label{plot:cwnd_32_wan_noting}}}
	\end{subfigure} 
	\begin{subfigure}[b]{\textwidth}
		\centering
		\includegraphics[scale=0.28]{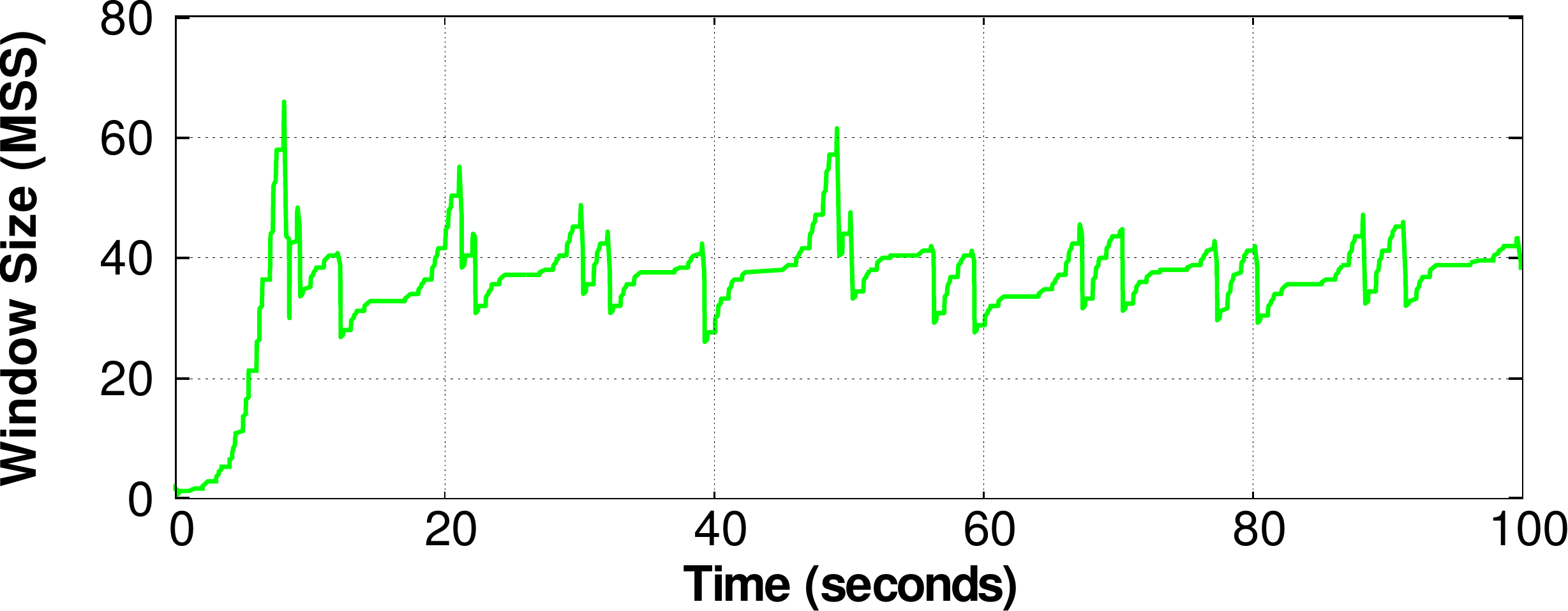}
		\caption{Without proxy, with time-slicing {\label{plot:cwnd_32_wan_ts}}}
	\end{subfigure}
	\caption{Congestion window behaviour of a LAN-WLAN TCP Cubic connection (experimental setup in Fig.~\ref{fig:lan_exp_setup}) when the \textbf{WLAN link is the bottleneck}: $C = 22\, Mbps$, $T_{on} = 360 \, ms$, $T_{off} = 640 \, ms$, $R_{in} = 32 \, Mbps$, and $RTPD = 150 \,ms$. \label{fig:cwnd_cubic_wan_32}}
\end{figure}

Now, if the WLAN link speed $R_{in}$ is increased to $32 \, Mbps$, the bottleneck shifts to the WLAN link and we expect the client to obtain a throughput of $22 \times \frac{360}{1000} = 8 \, Mbps$. The throughputs obtained by the client with time-slicing in the presence and absence of proxy for the $32 \, Mbps$ WAN link case are presented in Fig.~\ref{fig:throughput_cubic_wan_32}. The congestion window growth of a default (without proxy and without time-slicing) WAN-WLAN TCP connection is shown in Fig.~\ref{plot:cwnd_32_wan_noting}. By comparing  Fig.~\ref{plot:cwnd_32_wan_noting} and  Fig.~\ref{plot:cwnd_32_wan_ts}, we can see that the primary reason for the drop in the end-to-end throughput of time-sliced WAN-WLAN TCP, when the \emph{WLAN link is the bottleneck} and there is no proxy, is that the server's congestion window is unable to recover sufficiently during $T_{on}$.

\begin{figure}[ht]	
	\centering
	\begin{subfigure}[b]{\textwidth}
		\centering
		\includegraphics[scale=0.28]{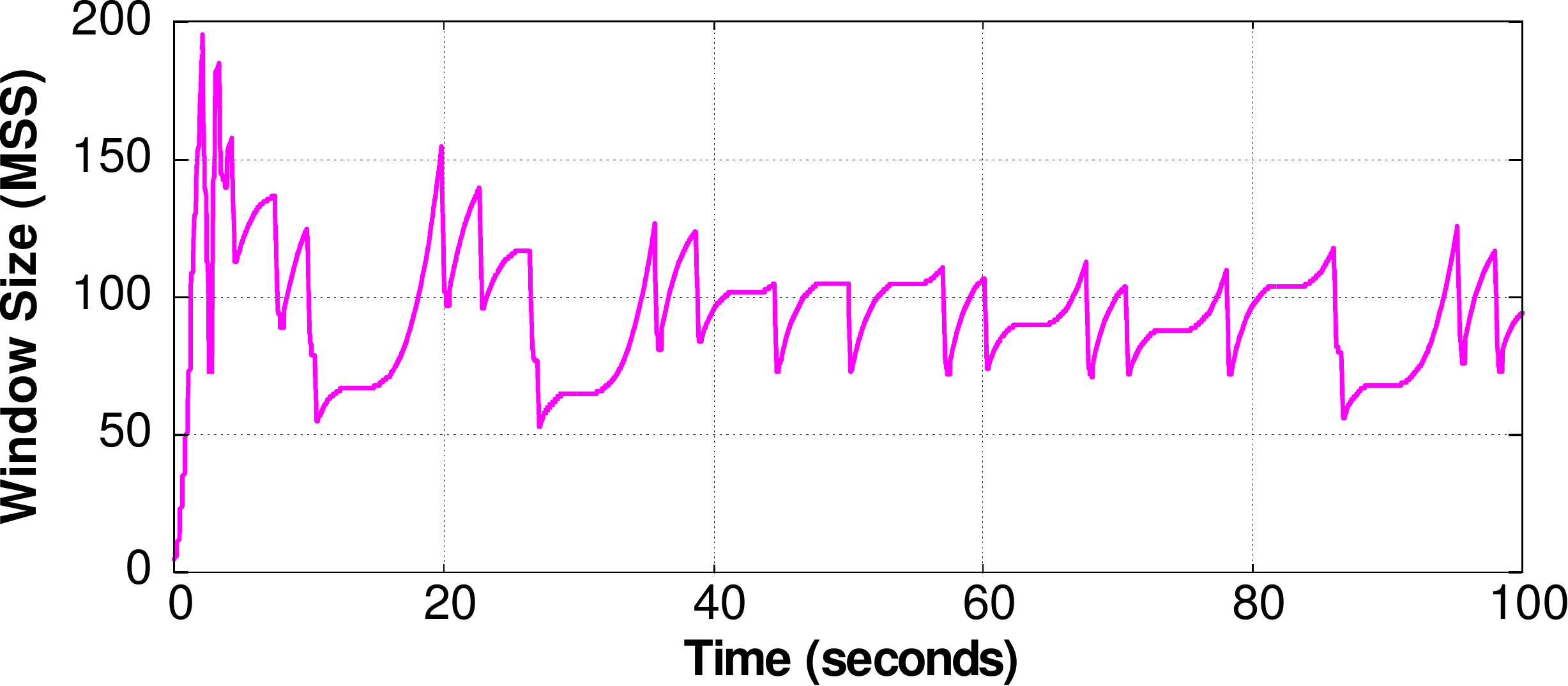}
		\caption{When the \textbf{WAN link is the bottleneck} : $R_{in} = 8 \, Mbps$. {\label{fig:cwnd_cubic_wan_8}}}
	\end{subfigure}
	\begin{subfigure}[b]{\textwidth}
		\centering
		\includegraphics[scale=0.28]{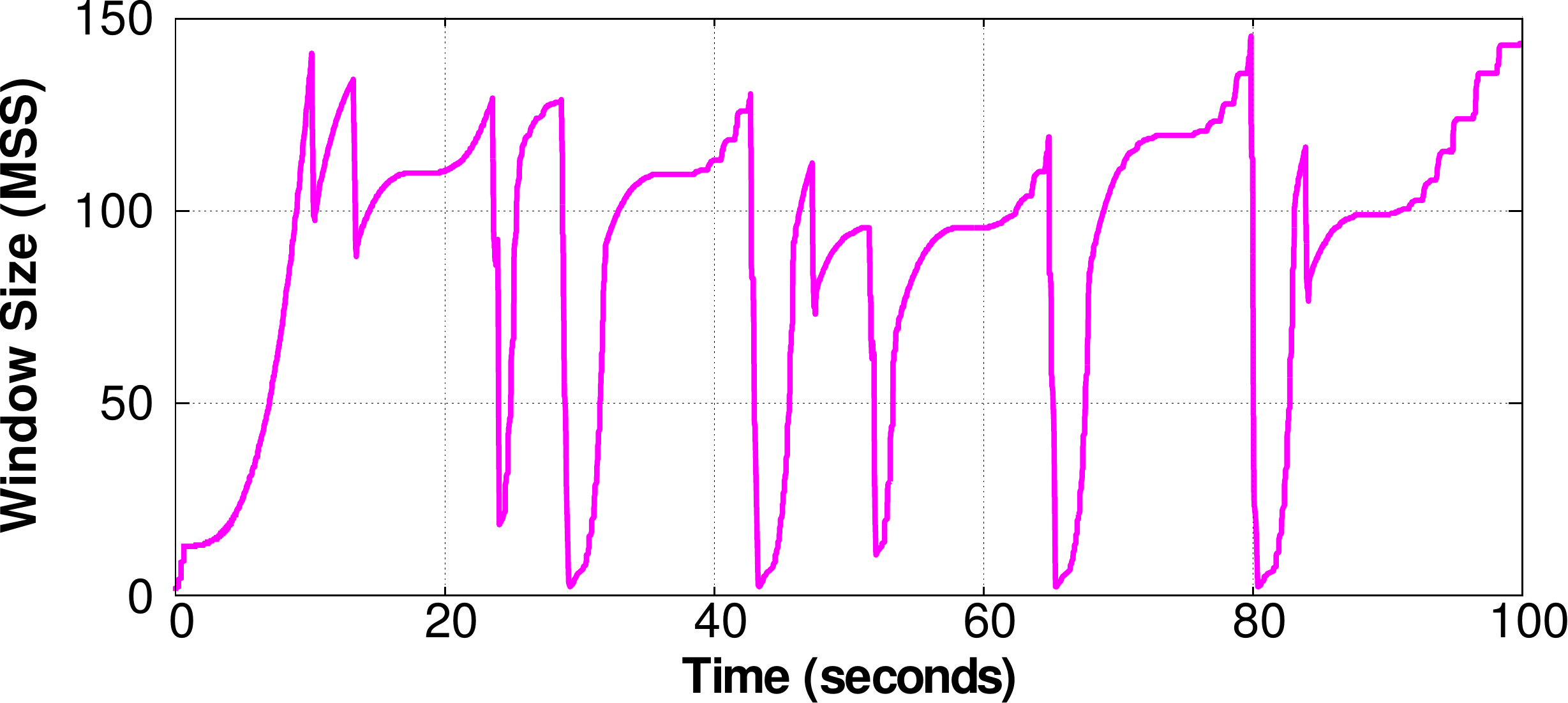}
		\caption{when the \textbf{WLAN link is the bottleneck}: $R_{in} = 32 \, Mbps$. {\label{plot:cwnd_32_wan_all}}}
	\end{subfigure}
	\caption{Congestion window behaviour of a LAN-WLAN TCP Cubic connection \textbf{with proxy and with time-slicing} (experimental setup in Fig.~\ref{fig:lan_exp_setup}): $C = 22\, Mbps$, $T_{on} = 360 \, ms$, $T_{off} = 640 \, ms$ and $RTPD = 150 \,ms$..}
\end{figure}

\begin{table}[!ht]
	\caption{Throughput of time-sliced WAN-LAN-WLAN TCP connection with proxy  when \textbf{WAN link is the bottleneck}; 
		the time-frame duration (T) is $1000 \, ms$, $C = 22 \, Mbps$, $\mathbf{R_{in} = 8 \, Mbps}$, RTPD is $150 \, ms$ and
		STA is associated at $54 \, Mbps$. \label{table:wan_table_8mbps_proxy}}
	\centering
	\begin{tabular}{|c|c|c|c|c|c|}
		\hline 
		\multirow{3}{*}{$T_{on}$} &  \multicolumn{4}{ c |}{Measured throughput (Mbps)} & \multirow{1}{*}{Expected}\\
		\cline{2-5}
		&  \multicolumn{2}{c|}{Cubic} & \multicolumn{2}{c|}{Reno} & \multirow{1}{*}{throughput} \\
		\cline{2-5}
		(ms) & F-RTO & No F-RTO & F-RTO & No F-RTO  & (Mbps)\\
		\hline 
		$80$   &  $1.93 \, $  & $1.91 \, $  &  $1.93 \, $  & $1.94 \, $ & $1.98 \, $\\
		\hline
		$180$   &  $3.92 \, $  & $3.91 \, $  &  $3.93 \, $  & $3.91 \, $ & $3.96 \, $\\
		\hline
		$290$   &  $6.36 \, $  & $6.33 \, $  &  $6.35 \, $  & $ 6.34 $ & $6.38 \, $\\
		\hline
		$360$   &  $7.92 \, $  & $7.92 \, $  &  $7.91 \, $  & $7.92 \, $ & $7.92 \, $\\
		\hline
	\end{tabular}
\end{table}

We notice that, unlike LAN-WLAN TCP connections, WAN-LAN-WLAN TCP connections are able to obtain only a fraction of the achievable throughput if they are time-sliced. However, when a TCP proxy is introduced between the WAN server and ADWISER~v2 (as in Fig.~\ref{fig:multi-ap-testbed}), data coming over the WAN link is cached by the proxy. This prevents the WLAN link from idling during $T_{off}$, in turn allowing the client to obtain the desired throughput of $8 \, Mbps$ (see the throughput plot ``with proxy, with time-slicing'' in Fig.~\ref{fig:throughput_cubic_wan_8}).  From Fig.~\ref{plot:cwnd_32_wan_all}, we can see that when the WLAN link is the bottleneck, the use of proxy allow the server's congestion window to recover during $T_{on}$. This in turn allows the client to achieve the expected throughput of $8 \, Mbps$, when the bottleneck shift to the WLAN link (see the throughput plot \emph{with proxy, with time-slicing} in Fig.~\ref{fig:throughput_cubic_wan_32}).  Also, when we have a $8 \, Mbps$ link (i.e., WAN is the bottleneck), the  congestion window of a TCP connection with \emph{time-slicing and with proxy} (Fig.~\ref{fig:cwnd_cubic_wan_8}) is similar to the congestion window of the default TCP Cubic.  To study the performance of time-sliced WAN-LAN-WLAN TCP transfers in the presence of a TCP proxy server, we performed long file downloads from the WAN server for various time slices. The results of our experiments (averaged over $10$ runs each) are presented in Tables~\ref{table:wan_table_8mbps_proxy} and \ref{table:wan_table_32mbps_proxy}.  From Tables~\ref{table:wan_table_8mbps_proxy} and ~\ref{table:wan_table_32mbps_proxy}, we can see that the measured
throughput is close to the expected throughput in all cases, as
the presence of proxy ``hides'' the large RTPD of the WAN.

\begin{table}[!ht]
	\caption{Throughput of time-sliced WAN-LAN-WLAN TCP connection with proxy when \textbf{WLAN link is the bottleneck}; 
		the time-frame duration (T) is $1000 \, ms$, $C = 22 \, Mbps$, $\mathbf{R_{in} = 32 \, Mbps}$, RTPD is $150 \, ms$ and
		STA associated at $54 \, Mbps$. \label{table:wan_table_32mbps_proxy}}
	\centering
	\begin{tabular}{|c|c|c|c|c|c|}
		\hline 
		\multirow{3}{*}{$T_{on}$} &  \multicolumn{4}{ c |}{Measured throughput (Mbps)} & \multirow{1}{*}{Expected}\\
		\cline{2-5}
		&  \multicolumn{2}{c|}{Cubic} & \multicolumn{2}{c|}{Reno} & \multirow{1}{*}{throughput} \\
		\cline{2-5}
		$(ms)$ & F-RTO & No F-RTO & F-RTO & No F-RTO  & (Mbps)\\
		\hline 
		$50$   &  $1.09 \, $  & $1.1 \, $  &  $1.15 \, $  & $1.16 \, $ & $1.10 \, $\\
		\hline
		$200$   &  $4.36 \, $  & $4.45 \, $  &  $4.39 \, $  & $4.41 \, $ & $4.40 \, $\\
		\hline
		$400$   &  $8.85 \, $  & $8.77 \, $  &  $8.82 \, $  & $8.83 \, $ & $8.80 \, $\\
		\hline
		$600$   &  $13.23 \, $  & $13.13 \, $  &  $13.25 \, $  & $13.15 \, $ & $13.20 \, $\\
		\hline
		$800$   &  $17.59 \, $  & $17.61 \, $  &  $17.58 \, $  & $17.59 \, $ & $17.60 \, $\\
		\hline
		$1000$   &  $22.00 \, $  & $22.00 \, $  &  $22.00 \, $  & $22.00 \, $ & $22.00 \, $\\
		\hline
	\end{tabular}
\end{table}

 The introduction of a TCP proxy raises the question: Is
it justified to assume the presence of a TCP proxy in the
real world? We note that enterprises often implement proxy
server, as shown in Fig.~\ref{fig:multi-ap-testbed}.  There are many reasons for installing a
proxy in a campus or enterprise setting: access control, access monitoring, and accounting being some of them. Another attractive feature of a proxy server is its ability to cache data. The ``Google Global Cache Service'' is essentially based on this idea \cite{ggc}. With a small number of GGC cache servers inside the network, GGC aims to quickly disseminate popular contents like YouTube videos to end users.
 
\section{Obtaining the Time Slices: Utility Optimization}
\label{sec:utility-optimization}

In this section, we formulate a constrained utility optimization problem whose solution yields the time slices. To do so, first, we capture the dependence among the links using a link dependence graph. Formally, we denote the link dependence graph as $G(\mathcal{V}, \mathcal{E})$, where $\mathcal{V}$ denotes the set of client-AP links in the network and $\mathcal{E}$ denotes the set of edges in graph $G$. For any two links $l_1, l_2 \in \mathcal{V}$,  edge $(l_1,l_2) \in \mathcal{E}$ if and only if transmissions from an endpoint of either of the links interferes with reception at an endpoint of the other link. 
Since we are dealing with TCP traffic where each end of a link has to serve as a transmitter and a receiver for any TCP connection on that link (due to TCP ACKs), we assume link dependence to be a symmetric relation. Therefore, the underlying dependence graph is undirected.

We recall from our discussion in Section \ref{sec:motivation} that scheduling dependent links leads to poor network performance. We therefore resort to the classical approach of independent set scheduling, i.e., scheduling dependent links in non-over lapping time slices. A subset of links $\mathcal{I} \subseteq \mathcal{V}$ in which no two links are dependent, and no other link can be added to the set $\mathcal{I}$ without resulting in a dependence is called a \emph{maximal independent set}. The collection of maximal independent sets $\mathbf{\mathcal{J}}$ in the network can be represented by a matrix $\mathbf{M}$, with entries ${M}_{jk}, j \in \mathcal{V}, k \in \mathbf{\mathcal{J}} $ as follows
$${M}_{jk} = 
\begin{cases}
1 & \textrm{if the link associated with client } j \textrm{ is in maximal independent set } k \\
0 & \textrm{otherwise}
\end{cases}
$$

It is well known that enumerating all the maximal independent sets in an arbitrary graph is a \emph{NP-hard} problem \cite{garey,nachiket}. Therefore, in typical networks, the size of matrix $\mathbf{M}$ could grow exponentially.  However, matrix $\mathbf{M}$ can still be computed for small networks.  For larger networks, as in \cite{nachiket}, we can use heuristics such as greedy link scheduling to solve the constrained optimization problem stated below. 

In ADWISER~v2, we have one virtual server and four virtual queues per client; a queue each for
WAN downloads, WAN uploads, LAN downloads and LAN
uploads. Segregating traffic in this way allows fine control
of service and  possible differentiation among various types
of traffic. Let $\eta_j$, $\xi_j$ and $\delta_j$ denote the non-negative weights assigned 
to WAN downloads, WAN uploads and LAN transfers (download and upload) of client $j$, 
respectively. We use a single weight for LAN transfers because
LAN downloads and uploads only use the WLAN resource and
we do not provide service differentiation between the two types of
LAN transfers; however, this can be easily extended.

Let $x_j$, $y_j$ and $z_j$ denote the fraction of time on the WLAN medium that is allotted to WAN downloads, WAN uploads and LAN transfers of client $j$, respectively. Then, the total fraction of time the WLAN medium is allocated to client $j$ is given as $(x_j + y_j + z_j)$.
Let $v_j$ be the service rate of the virtual server
corresponding to client $j$. Then, our optimization problem can be stated as
follows:
\begin{align}
\max_{\mathbf{x}, \mathbf{y}, \mathbf{z}, \mathbf{a}} \Big(  \sum_{j \in \mathcal{N}} \eta_j   U ( x_j  \cdot v_j )  +  \xi_j   U
&  (y_j \cdot v_j ) +  \delta_j  U (z_j \cdot v_j ) \Big) & \nonumber \\
\textrm{Subject to: } \hspace{0.3\linewidth} &  \nonumber \\
\sum_{j \in \mathcal{N}} \left( x_j + \alpha  y_j \right) \cdot v_j \leq r_{i}\,  ,  \, \, \,  \sum_{j \in \mathcal{N}} & \left( \alpha  x_j + y_j \right)  \cdot v_j \leq r_{o} \label{eq:multi_ap_wan_constraint} \\
\left( x_j + y_j + z_j \right)  \leq \sum_{k \in \mathcal{J}}  M_{jk} &  \cdot a_k \, , \quad  \sum_{k \in \mathcal{J}} a_k  \leq 1  \label{eq:capacity_constrint} 
\end{align}
\begin{align}
\sum_{j \in \mathcal{N}} ( x_j + y_j +  z_j )  & \leq 1 \label{eq:set1}\\ 
\mathbf{x}   \geq  \mathbf{0}, \, \mathbf{y} \geq \mathbf{0},  \, \mathbf{z}  \geq \mathbf{0}, & \, \mathbf{a} \geq \mathbf{0}\label{eq:set2} 
\end{align} 
where $r_{i}$ and $r_{o}$ are the inbound and outbound bit rates of the WAN access link, and  $U(\cdot)$ is a strictly concave twice differentiable increasing function, respectively. Such assumptions on the utility functions are very standard. For example, $\log⁡(\cdot)$ is a popular and well-studied concave utility function. The assumption of concavity can be attributed to the law of diminishing marginal returns. Also, the concavity assumption makes the optimization problem more tractable. In the above problem formulation, $a_k$ represents the fraction of time the independent set $k$ is active. Constraints \eqref{eq:capacity_constrint} and \eqref{eq:multi_ap_wan_constraint} represent the WLAN and WAN capacity constraints, respectively. In the inequality constraint
\eqref{eq:multi_ap_wan_constraint}, $\alpha \in (0,1)$
represents the ``ACK loading factor,'' which is the ratio of TCP ACK bits transferred in the opposite direction for bits corresponding to a data packet transfer. In constraint \eqref{eq:multi_ap_wan_constraint}, $\alpha y_j v_j$ and $\alpha x_j v_j$ correspond to the inbound and outbound ACK rates, respectively. We have obtained $\alpha$ as follows.
Consider a TCP download over the WLAN. For every two data packets ($\sim 3 \, KBytes$), there will be one ACK packet ($52 \, Bytes$) in the uplink direction (assuming delayed ACK behaviour). Thus, $\alpha$ equals  $\frac{52}{3000}$.   Since the optimization problem stated above maximizes a concave function subject to linear constraints over a convex set,  it can be solved using techniques from convex optimization \cite{Boyd}. Let the tuple $(\mathbf{x}^{*}, \mathbf{y}^{*}, \mathbf{z}^{*}, \mathbf{a}^{*})$ be an optimizer of the above problem. Then, we can obtain a schedule as follows. In every time-frame of duration $T$, for each maximal independent set $k \in \mathcal{J}$, allow transmission to the clients in the set $k$ for a duration of $a^{*}_k \cdot T$ units of time. We note that $T$ should be large enough so that the TCP transients die out in a small fraction of $T$.  However, if the value of $T$ is too large, it can affect the performance of short-lived and interactive TCP transfers. It has been our experience over several experiments that, for the experiments presented in this paper, a value of $1000 \, ms$ is ideal for $T$. Since the value of $T$ affects the performance of short-lived and interactive TCP traffic, a more comprehensive mechanism is needed to determine $T$ such as incorporating $T$ in the utility function. We plan to pursue this in our future work. 

\section{Inferring Link Dependencies and Rate Adaptation for Long-lived TCP transfers}
\label{ref:dld_ra}

Until now, we have assumed that the physical rate of association of the
STAs and the link dependencies are available to us \textit{a priori} 
and are time-invariant. 
However, in reality, they are arbitrary and vary over time. Therefore, we need to be able to infer link dependencies dynamically
and adapt the service rate of the virtual servers, to effectively utilize the
wireless medium.

\subsection{Inferring Dependence: A client-assisted Technique} 
\label{sec:dynamic_link_dependency}
In this section, we propose an online low-overhead heuristic to infer link dependencies in the network.
When two clients are associated with the same AP, we declare the corresponding links to be \emph{dependent}. 
Now, consider two clients $S_1$ and $S_2$. Let clients $S_1$
and $S_2$ be associated with access points $A_1$ and $A_2$,
respectively. We classify the dependence between the clients into
the following three types. 

\begin{figure}[ht]
	\centering
	\begin{tabular}{ccc}
		\begin{subfigure}[b]{0.3\textwidth}
			\centering
			\includegraphics[scale=0.3]{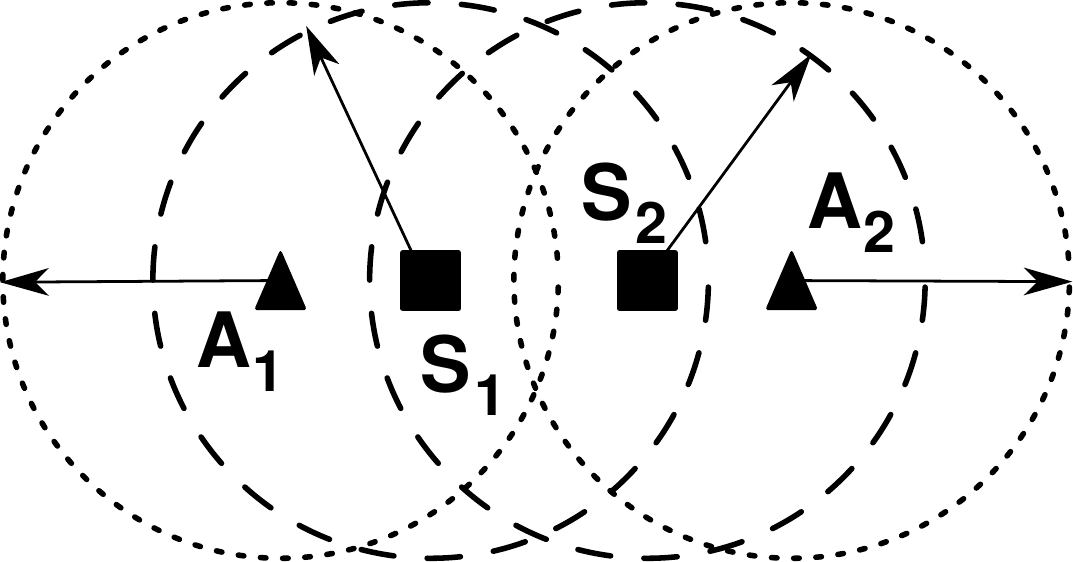}
			\caption{Type I dependence}
			\label{fig:dep1}
		\end{subfigure}
		&
		\begin{subfigure}[b]{0.3\textwidth}
			\centering
			\includegraphics[scale=0.3]{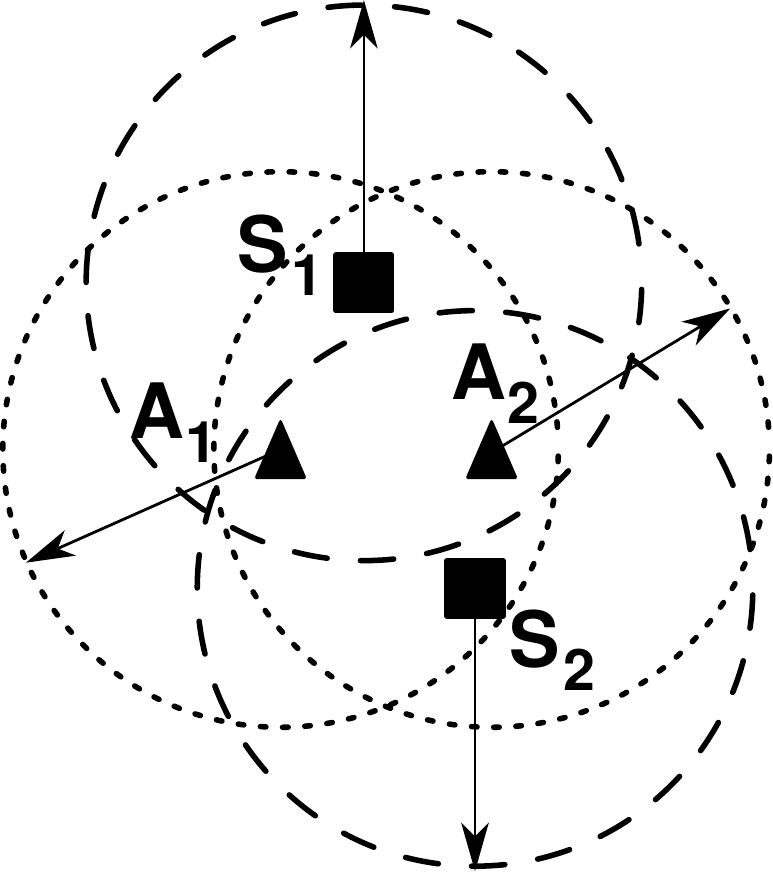}
			\caption{Type II dependence}
			\label{fig:dep2}
		\end{subfigure} 
		&
		\begin{subfigure}[b]{0.3\textwidth}
			\centering
			\includegraphics[scale=0.3]{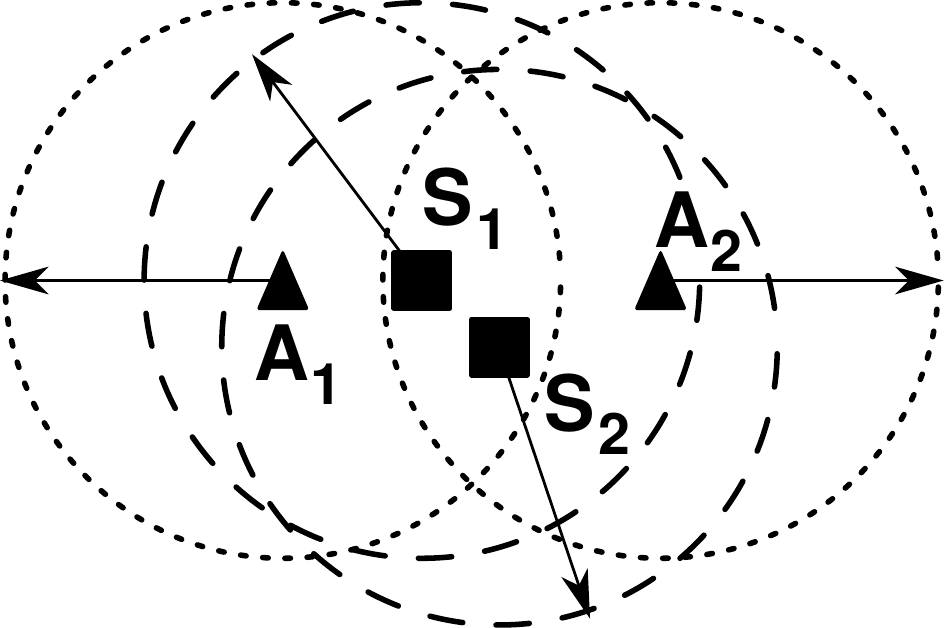}
			\caption{Type III dependence}
			\label{fig:dep3}
		\end{subfigure}  \\ 
	\end{tabular} 	
	\caption{Figures illustrating the various types of link dependencies in WLANs. The dotted and dashed circles denote the interference ranges of APs and clients, respectively. The APs and clients are represented as darkened triangles and squares, respectively.}
\end{figure}

\begin{itemize}
	\item \textit{\textbf{{Type \Rmnum{1}} Dependence}} (Fig.~\ref{fig:dep1}): Clients $S_1$
	and $S_2$ are within interference range of each other, and the clients
	are outside the interference ranges of each other's access
	points. Further, the access points $A_1$ and $A_2$ do not interfere with
	each other.
	\item \textit{\textbf{{Type \Rmnum{2}} Dependence}} (Fig.~\ref{fig:dep2}): Access points
	$A_1$ and $A_2$ interfere with each other.
	\item \textit{\textbf{{Type \Rmnum{3}} Dependence}} (Fig.~\ref{fig:dep3}): Access points
	$A_1$ and $A_2$ do not interfere with each other. Client $S_2$ is
	within the interference range of access point $A_1$ or client
	$S_1$ is within the interference range of access point $A_2$.
\end{itemize}

\emph{Type \Rmnum{1} dependencies} are difficult. Therefore, as in \cite{survey.shrivastava-etal09centaur}, we too ignore them. In networks with dense AP deployment, ignoring  \emph{Type \Rmnum{1} dependencies} will not impact the network performance because they have a low probability of occurrence (Table \Rmnum{1} in \cite{prob_analysis_albert}). \emph{Type \Rmnum{2} Dependencies} are time invariant, and depend only on the location of the APs. They can be evaluated after deploying the APs, and stored in ADWISER~v2.  Next, we provide a heuristic to infer \emph{Type \Rmnum{3} dependencies} in the network. 
Consider a network of $\mathcal{N}=\{1,2,\cdots,n\}$ clients and $\mathcal{M}=\{1, 2, \cdots, m\}$ co-channel APs.
For each client $j \in \mathcal{N}$, let $m_j$ denote the AP with which client $j$ is associated. 
Also, for each client $j \in \mathcal{N}$, let $P_{k,j}$ denote the received signal strength of the beacon of
access point $k \in \mathcal{M}$, reported by client $j$. Let $\mathcal{N}_k \subseteq \mathcal{N}$ denote the set of clients associated with AP $k$. Then, the \emph{Type \Rmnum{3} dependencies} can be dynamically discovered  using the following heuristic.

\begin{algorithm}[ht]
	\caption{Heuristic for inferring  \emph{Type \Rmnum{3} dependencies} \label{algo:link_infer}}
	\begin{algorithmic}[1]
		\FOR{each $j \in \mathcal{N}$ and $k \in \mathcal{M}$}
		\IF{client $j$ and AP $k$ are on the same channel}
		\IF{$\frac{P_{m_j,j}}{P_{k , j}} < p_{th}$ }
		\FOR{each $i \in \mathcal{N}_k$}
		\STATE clients $i$ and $j$ are pair-wise dependent
		\ENDFOR
		\ENDIF
		\ENDIF
		\ENDFOR
	\end{algorithmic}
	\label{algo:dld}
\end{algorithm}

Algorithm~\ref{algo:link_infer} uses the relative beacon strength to declare dependencies. To illustrate the working of Algorithm~\ref{algo:link_infer}, let us consider two clients $S_1$ and $S_2$ associated with access points $A_1$ and $A_2$, respectively.  Let the received signal strength reported by client $S_1$ be $P_{1,1}$ and $P_{2,1}$ (corresponding to access points $A_1$ and $A_2$, respectively). Similarly, client $S_2$ also reports received signal strength $P_{1,2}$ and $P_{2,2}$ (corresponding to access points $A_1$ and $A_2$, respectively).  Algorithm~\ref{algo:dld} declares clients $S_1$ and $S_2$ as dependent if $\frac{P_{1,1}}{P_{2,1}} < p_{th}$ or $\frac{P_{2,2}}{P_{1,2}} < p_{th}$. Since Algorithm~\ref{algo:link_infer} uses the relative beacon strength to infer dependencies, it can infer dependencies even when the client are associated at different physical rates.

The threshold $p_{th}$ depends on the environment and can be found, experimentally, using the following methodology. Associate two clients $S_1$ and $S_2$ with two non-interfering APs $A_1$ and $A_2$, respectively. Place the clients at various locations. Each location corresponding to a scenario. For each scenario, note down the throughput and received signal strength of the clients during standalone and simultaneous long-lived TCP downloads. By comparing the throughputs during standalone and simultaneous transmission, we can conclude if the clients are dependent, i.e., interfering with each other. When the clients do not interfere with each other, we compute the threshold value as $\min \left\{ \frac{P_{1,1}}{P_{2,1}} , \frac{P_{2,2}}{P_{1,2}} \right\}$.  The threshold $p_{th}$ is obtained as the average of threshold value of several scenarios.

To facilitate the discovery of link dependencies, we run a lightweight application level program at each client. This program periodically scans the channel on which its host is  associated. During the scan, the program measures the received signal strength of the beacon from \emph{all the co-channel access points.} When the scan is complete,  the program report the scan results to ADWISER~v2 over UDP.  Upon reception of the results, ADWISER~v2 checks all the pair-wise dependence criteria and updates the dependence graph accordingly. The scanning operation generates a packet (of size few kilobytes) per second at each client. So, even in dense networks, the overhead due to the scan reports is minuscule. Further, in dense networks, we can reduce the frequency of scans, thereby reducing the overhead. \emph{We would like to remark that these scan results are the only form of communication between the clients and ADWISER~v2, and  takes place over UDP --- this does not require any modifications to the firmware or hardware of the clients and APs.}

\subsection{Adaptive Estimation of Service Rate for Long-lived TCP Transfers} 
\label{sec:rate_adaptation}

The time-slicing algorithm dictates that if client $j \in \mathcal{M}$ is
scheduled in the $t^{\textrm{th}}$ time slice, then the packets in its virtual queue need to be
served at a constant rate of
$v_j(t)$ for the duration of the time slice. If we set $v_j(t)$ to a low value, the virtual server will become the bottleneck, resulting in inefficient utilization of the WLAN medium.
Setting $v_j(t)$ to a high value may end up shifting the
queue to the access point. Keeping the queues in ADWISER~v2 permits us to control the release of packets. If the queues move to the APs, then we lose control over them, and ADWISER~v2 can no longer manage the TCP flows. This will result in
poor and unpredictable throughputs (seen unmanaged mode of operation in Section~\ref{sec:motivation} ). Thus, we need to set
$v_j(t)$ to a value slightly lower than  $\mu_j$, where $\mu_j$ is the average TCP rate of client $j$ over the wireless medium. However, $\mu_j$ is a function of various randomly-changing factors like distance from the access
point and the environment, and is not known \emph{a priori}. 

\begin{figure}[ht]
	\centering
	\includegraphics[scale=0.45]{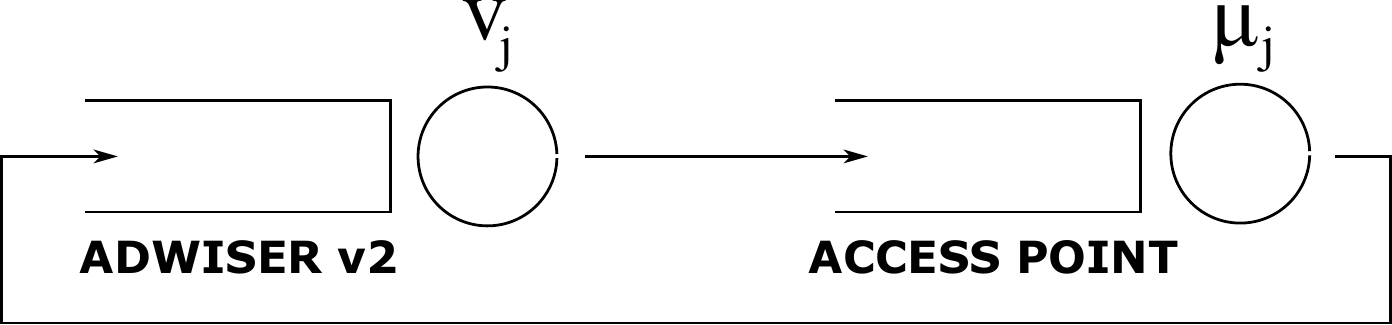}
	\caption{A closed queuing network representing TCP connection in steady state.}
	\label{fig:tcp_steady_state}
\end{figure}

Due to the shielding effect of the TCP proxy and the isolation of flows due to
time slicing, at steady state, each TCP connection can be modelled by
the closed queuing network shown in Fig. \ref{fig:tcp_steady_state}.
Let $q_j(t)$ denote the queue length (in $KBytes$) immediately after service of 
the queue corresponding to client $j$. Then, the rate adaptation algorithm at the virtual server in ADWISER~v2 for client $j \in \mathcal{N}$ is as described by Algorithm \ref{algo:rate_adapt}.

\begin{algorithm}[h]
	\caption{Rate Adaptation Algorithm for client $j \in \mathcal{N}$}
	\begin{algorithmic}[1]
		\STATE Set $\overline{q}_j(t) := 0, \, v_j(0) := 1 \, Mbps, t_{prev} = 0$
		\WHILE{ADWISER~v2 is in managed mode}
		\IF{end of current time-frame $ \geq (t_{prev} + t_{interval})$}
		\IF{client $j$ was scheduled in time slice $t$}
		\IF{$\overline{q}_j(t) > q^{lb}_j$}
		\STATE $v_j(t+1) :=  v_j(t)+  \epsilon \cdot \overline{q}_j(t)$
		\ELSE
		\STATE $v_j(t+1) :=  \lambda \cdot v_j(t)$
		\ENDIF
		\STATE $\overline{q}_j(t+1) :=  \beta \cdot {q}_j(t) + (1 - \beta) \cdot \overline{q}_j(t)$
		\STATE $t_{prev} := t_{prev}+t_{interval}$
		\ENDIF
		\ENDIF
		\ENDWHILE
	\end{algorithmic}
	\label{algo:rate_adapt}
\end{algorithm}

In Algorithm	~\ref{algo:rate_adapt}, $\epsilon > 0$, $\lambda \in (0,1)$ and $\beta \in [0,1]$ are the control parameters.   The rate adaptation algorithm runs every $t_{interval}$ units of time.  Algorithm \ref{algo:rate_adapt} adapts the service rate of the virtual server based on the average queue length $\overline{q}_j$. If $\overline{q}_j$ falls below $q^{lb}_j$, then the queue has a tendency to shift to the access point. Therefore, in such cases, the best course of action is to reduce the service rate of the queue. On the other hand, if the $\overline{q}_j$ starts to increase, then the implication is that the current service rate is less than the capacity offered by the wireless medium. Under such circumstances, we increase the service rate in proportion to the average queue length. The parameter $\beta$ can be used to control the rate of convergence of the algorithm. If $\beta$ is large, then less importance is given to the history of the queue, and the algorithm will react quickly to changes in queue length. This, in turn, would lead to fluctuations in the throughput of the client.  On the other hand, if $\beta$ is small, then the average queue length is biased towards the past, and the algorithm will not react quickly to changes in queue length.

\section{Experimental Results}
\label{sec:experimental_results}

The experimental results presented in this section demonstrate that coarse time-slicing can meet the utility optimization objective discussed in Section~\ref{sec:utility-optimization}. The experiments reported in this paper were conducted in a busy academic building with IEEE 802.11 WLAN. We would like to mention that, in all our experiments, the hardware/firmware of the APs and the clients in the network were not modified in any way. For ease of presentation, we have restricted ourselves to topologies for which the expected performance could be easily computed. While many of the experiments discussed subsequently are performed on  IEEE~802.11g infrastructure WLANs, the issues we consider remain relevant in IEEE~802.11n infrastructure WLANs, as well as other newer IEEE~802.11 WLAN variants.

\subsection{The Testbed}
\label{sec:testbed}

\begin{figure}[ht]
\centering
  \includegraphics[scale=0.17]{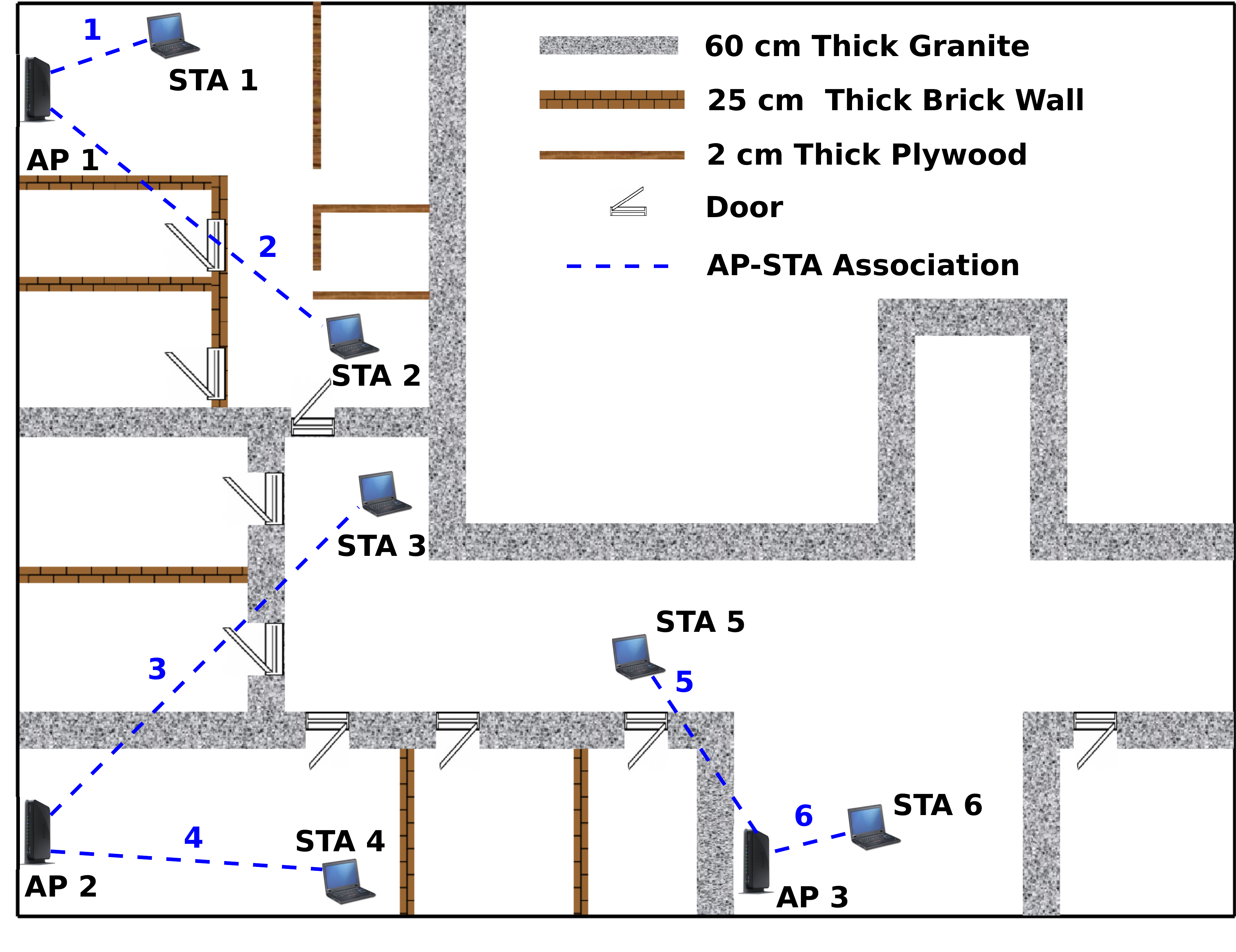}
  \caption{The layout of the ADWISER~v2 testbed. There are 3 IEEE~802.11g APs and 6 clients,
    all on Channel~11. The dashed lines indicate client-AP associations. The composition of the walls is shown in the legend.}
  \label{fig:ece_layout_for_paper}
\end{figure}

ADWISER~v2 has been developed on a Fedora Linux platform. The system runs on a 1U rack-mountable Quad core Intel
Xeon 3.2GHz system. Apart from ADWISER~v2, the testbed consists of various components as described
here. We used a Fedora Linux based server (Quad-core
Intel Xeon E5620 2.40 GHz processor, 4 GB with DDR-3 1333 MHz RAM) and laptops
(Intel Core i5 2.40 GHz processor, 4 GB with DDR-3 1333 MHz RAM) as end
systems. Atheros chipset based IEEE~802.11g Netgear WNDR3700v2 APs, flashed
with DD-WRT supporting the popular MadWiFi wireless stack, have been deployed
as part of the testbed infrastructure. The default \emph{Minstrel} rate
adaptation algorithm is used by the APs. Further, the 
RTS/CTS mechanism is disabled in all the APs because this mechanism could bring down the throughput by as much as $50\%$ \cite{cheng}.  A commercial router was used for creating the Internet access link, and the
Linux \emph{netem} utility employed to implement WAN propagation delays. The
Fedora Linux Squid-3.1 has been employed as a proxy
server (Intel Core i5 2.40 GHz processor, 4 GB with DDR-3 1333 MHz RAM).
 Custom scripts, along with \emph{wget}
and \emph{iperf}, have been used for traffic generation. Further, in our experiments, we have set all the weights mentioned in
Section~\ref{sec:utility-optimization} to unity. The values of the various
parameters of the link inference heuristic, and rate adaptation algorithm used in our experiments are as follows: $t_{interval}=3 \,seconds$, 
$q_j^{lb}=7.5 \, KBytes$, $\epsilon=0.05 \, Mbps$, $\lambda=0.9$, $\beta=0.8$ and $p_{th} = 0.3$, and $U(\cdot)$ is chosen as $log(\cdot)$.

\subsection{3 APs, 6 clients; LAN-WLAN Transfers} \label{sec:test_case_3}

\begin{figure}[ht]
	\centering
	\includegraphics[scale=0.35]{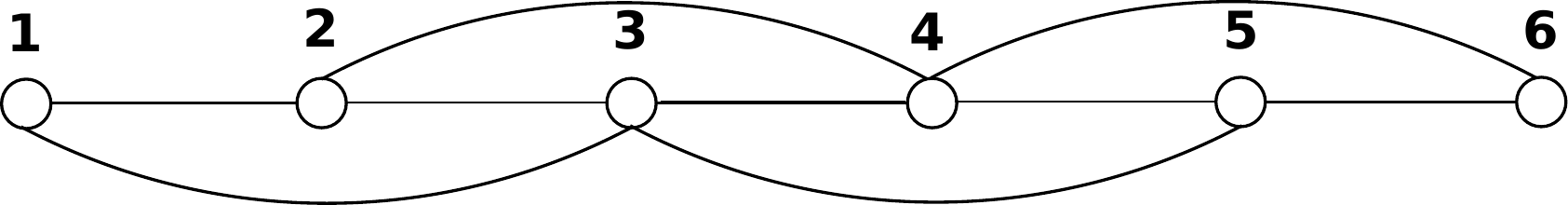}
	\caption{Dependence graph for the network in Fig.~\ref{fig:ece_layout_for_paper}.}
	\label{fig:3ap_6sta_dependence_graph}
\end{figure}

\begin{figure}[ht]
	\centering
	\includegraphics[scale=0.4]{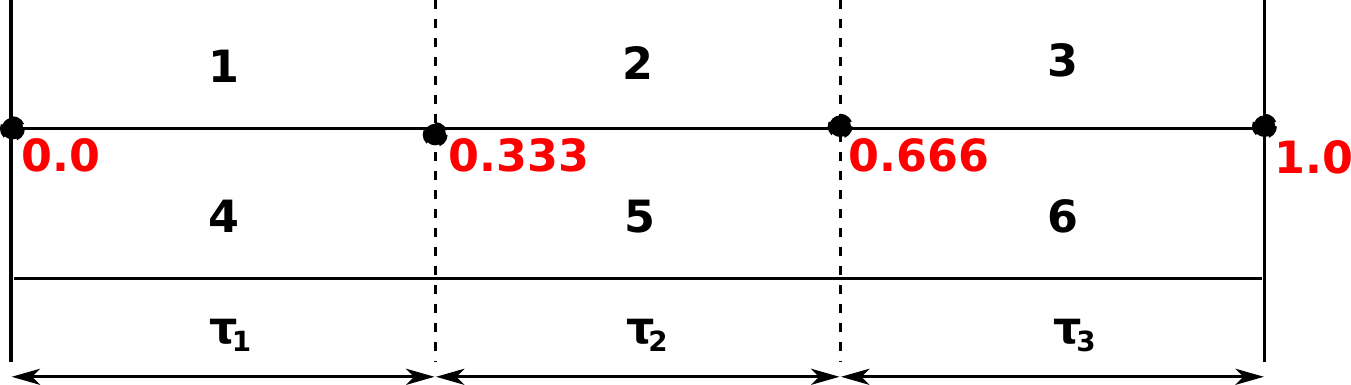}
	\caption{3 APs, 6 clients; LAN-WLAN transfers (experimental setup in Fig.~\ref{fig:ece_layout_for_paper}):   Allocation of clients within a time-frame of duration $1000 \, ms$.}
	\label{fig:six_sta_all_LAN_2_timeslices}
\end{figure}

This experiment demonstrates how ADWISER~v2 can mitigate interference in multi-AP WLANs.
The link dependence graph for this scenario is shown in Fig.~\ref{fig:3ap_6sta_dependence_graph}. We have 6 clients  downloading large files from a server on the LAN. The solution of the optimization
problem yields the time slices. These are
depicted in Fig.~\ref{fig:six_sta_all_LAN_2_timeslices}. 

\begin{figure}[ht]
		\centering
		\includegraphics[scale=0.33,angle=-90]{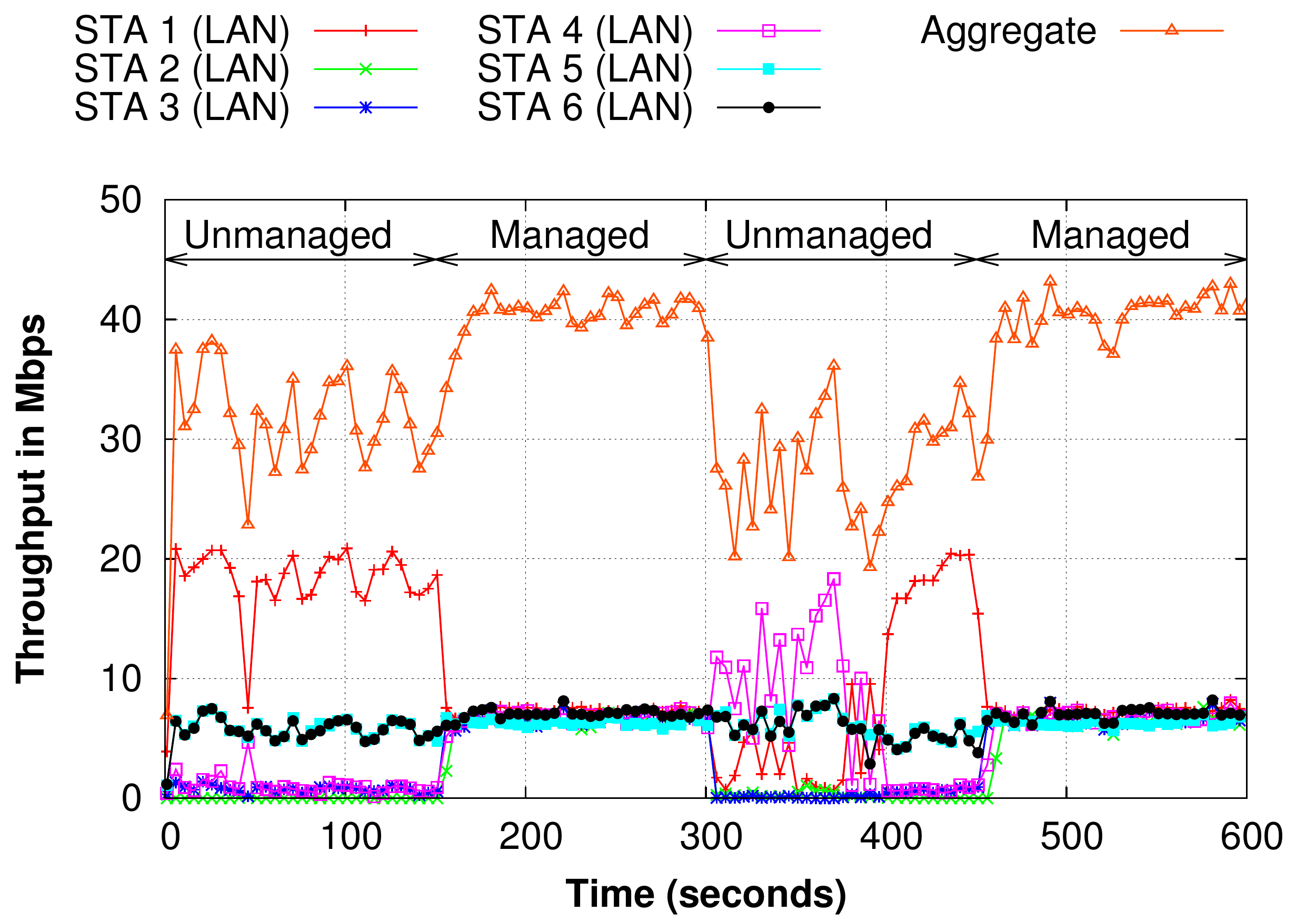}
		\caption{3 APs, 6 clients; LAN-WLAN transfers (experimental setup in Fig.~\ref{fig:ece_layout_for_paper}): Individual and aggregate throughputs obtained by the clients, in the unmanaged and managed modes.
			\label{plot:multi_ap_six_sta_exp_2}}
\end{figure}	
\begin{figure}[ht]
	\centering
	\includegraphics[scale=0.35]{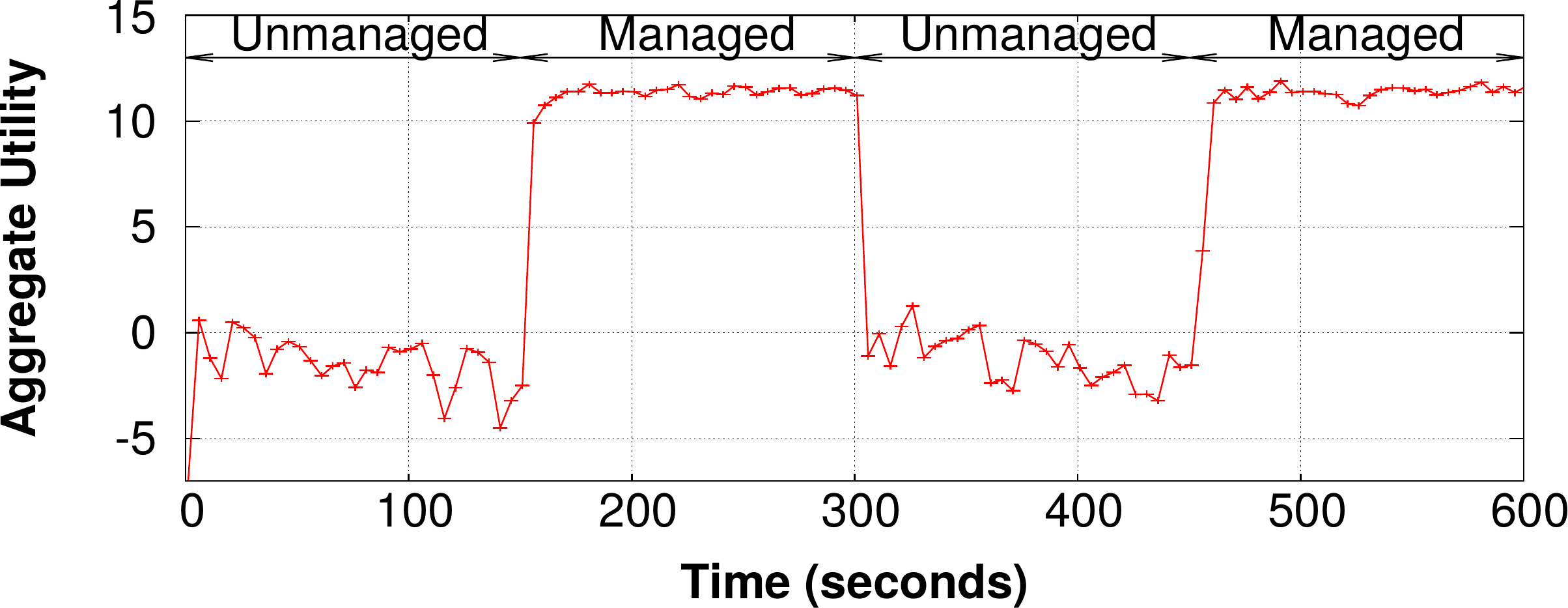}
	\caption{3 APs, 6 clients; LAN-WLAN transfers (experimental setup in Fig.~\ref{fig:ece_layout_for_paper}): Aggregate utility in the unmanaged and managed modes.}
	\label{fig:3ap_6sta_all_lan_utility}
\end{figure}

Fig.~\ref{plot:multi_ap_six_sta_exp_2} shows the throughputs of the 
STAs, in the managed and unmanaged modes. In the unmanaged mode, STA~1 alone gets a large throughput between $15 \, Mbps$ and
$20 \, Mbps$; STA~2, STA~3 and STA~4 get low throughputs, with STA~2 getting
almost nothing. With STA~3 and STA~4 ``suppressed,'' STA~5 and STA~6 obtain
throughputs between $5 \, Mbps$ and $7 \, Mbps$. The aggregate throughput varies
between $20 \, Mbps$ and $37 \, Mbps$. When the network is managed by ADWISER~v2, only independent clients are
scheduled, and all of them obtain fairly flat throughputs of about
$7 \, Mbps$, yielding a slightly variable aggregate throughput with an
average of about $41 \, Mbps$. The improvement in aggregate utility 
due to fair allocation of WLAN resources is shown in Fig. \ref{fig:3ap_6sta_all_lan_utility}.

\subsection{2 APs, 4 clients; LAN-WLAN and WAN-WLAN Transfers} \label{sec:test_case_4}

\begin{figure}[ht]
	\centering
	\includegraphics[scale=0.4]{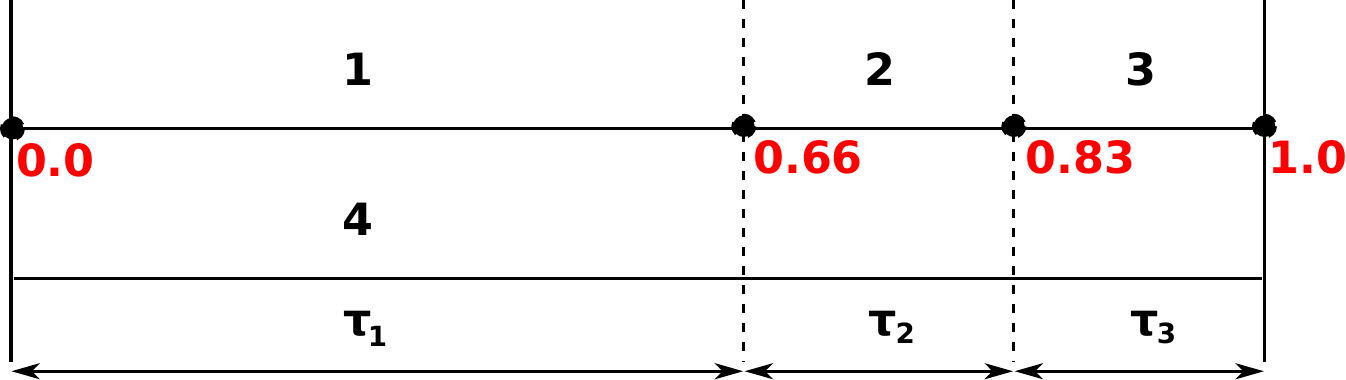}
	\caption{2 APs, 4 clients, LAN-WLAN and WAN-WLAN transfers:  Allocation of clients within a time-frame of duration $1000 \, ms$.}
	\label{fig:four_sta_exp_2_time_slicing}
\end{figure}

This experiment demonstrates the LAN-WAN fairness provided by ADWISER~v2 in multi-AP WLANs.
The physical position of the clients are shown in Fig.~\ref{fig:ece_layout_for_paper} . STA~2 and STA~3 are downloading large files from the WAN server, whereas STA~1 and STA~4 are downloading large files from the LAN server. There is a
round-trip propagation delay of $150 \, ms$ over the emulated Internet,
with the access link being $8 \, Mbps$. The solution of the
optimization problem yields the time slices and the sets of clients served in each time slice 
(see Fig.~\ref{fig:four_sta_exp_2_time_slicing}).

\begin{figure}[ht]
	\centering
	\includegraphics[scale=0.33,angle=-90]{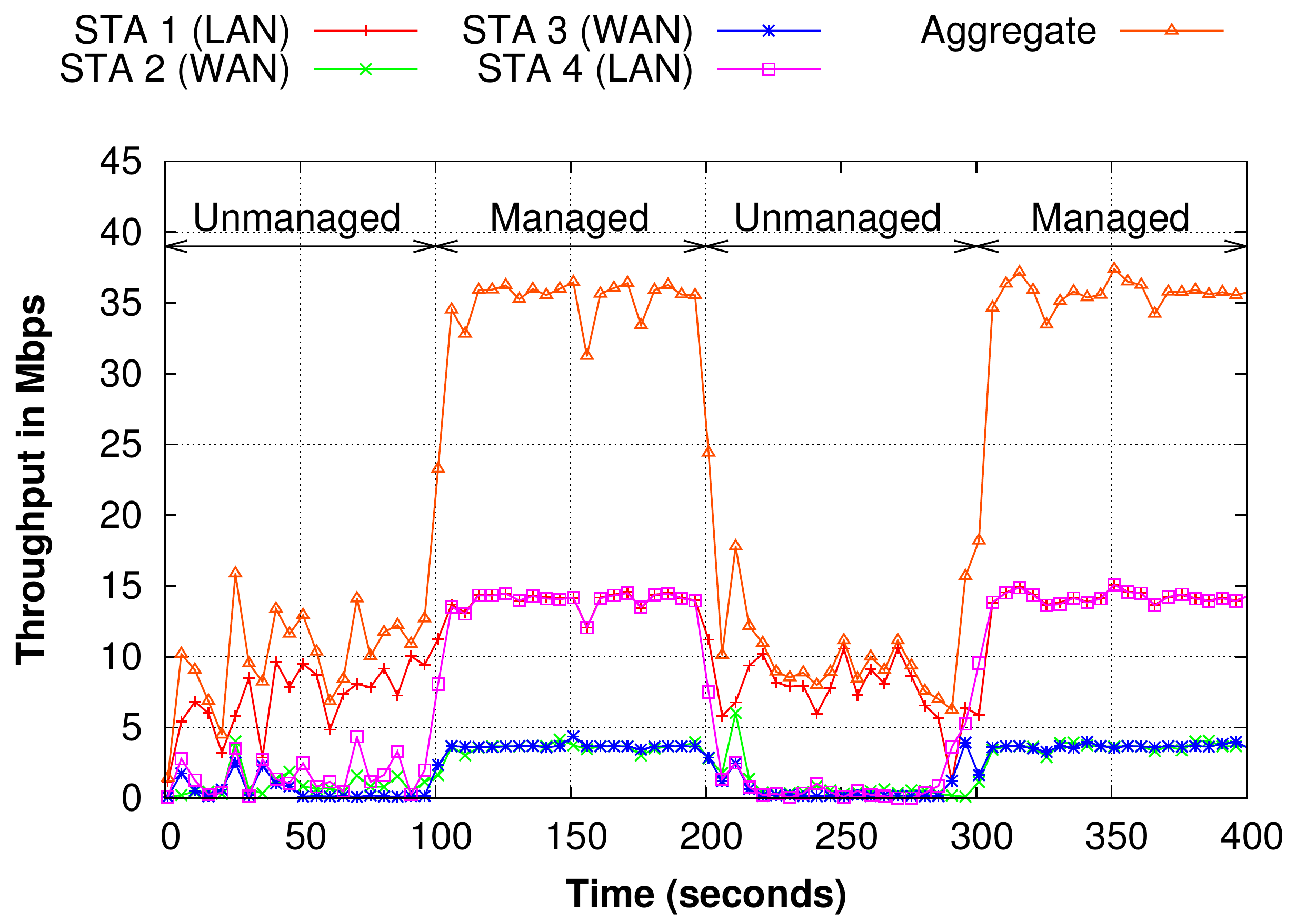}
	\caption{2 APs, 4 clients; LAN-WLAN and WAN-WLAN transfers:  Individual and aggregate throughputs obtained by the clients, in the unmanaged and managed modes.
		\label{plot:multi_ap_four_sta_mixed_exp}}
\end{figure}

\begin{figure}[ht]
	\centering
	\includegraphics[scale=0.35]{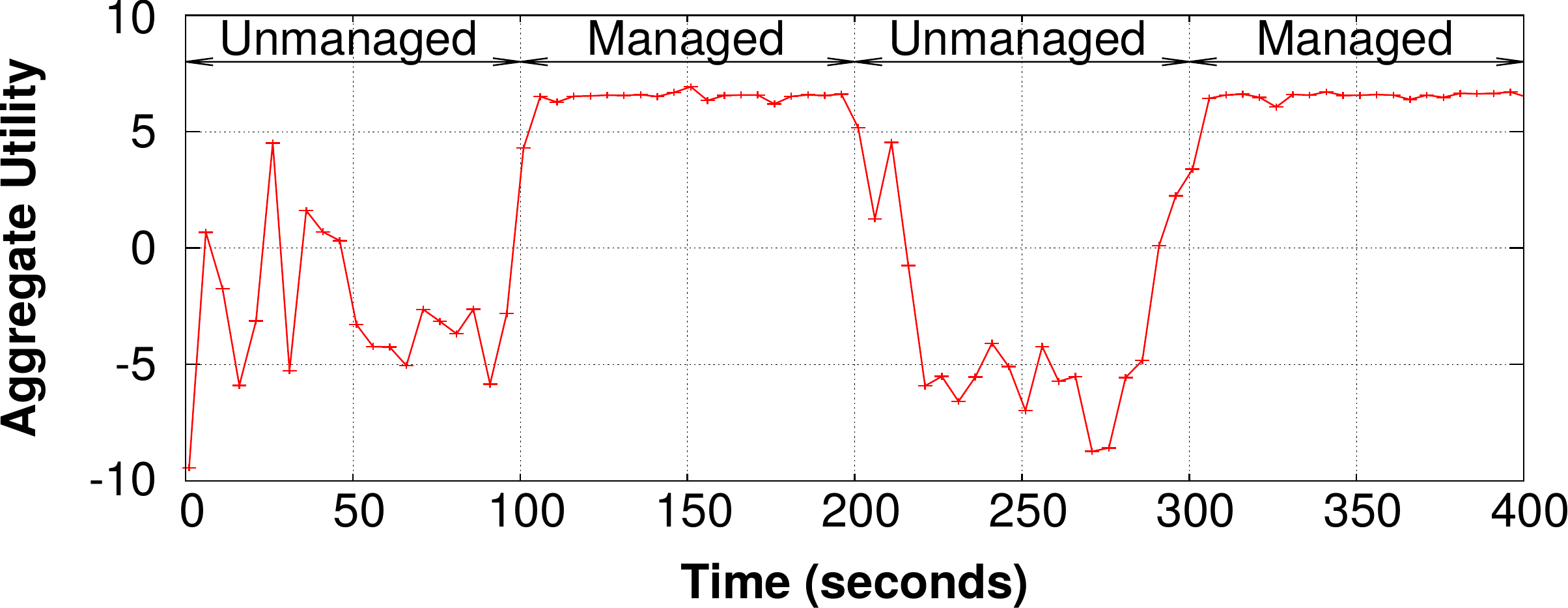}
	\caption{2 APs, 4 clients; LAN-WLAN and WAN-WLAN transfers: Aggregate utility in unmanaged and managed modes.}
	\label{fig:2ap_4sta_wan_lan_utility}
\end{figure}

Fig.~\ref{plot:multi_ap_four_sta_mixed_exp} shows the throughputs 
of the clients, in the managed and
unmanaged modes. In the unmanaged mode, the observations are similar to the previous experiments. 
In the managed mode, STA~2 and STA~3 obtain $3.7 \, Mbps$ each. The remaining time on the WLAN medium
is used by STA~1 and STA~4 \emph{concurrently}, each of them obtaining about $14 \, Mbps$. 
The aggregate throughput in the manged mode is over $35 \, Mbps$; substantially more than that the aggregate throughput of about $10 \, Mbps$ in the unmanaged mode. The improvement in aggregate utility obtained in managed mode is due to 
fairness among flows as shown in Fig.~\ref{fig:2ap_4sta_wan_lan_utility}.

\subsection{2 APs, 4 Mobile Clients; LAN-WLAN Transfers} \label{sec:test_case_5}

\begin{figure}[!ht]
	\centering	
	\begin{subfigure}[b]{0.45\textwidth}
		\centering
		\includegraphics[height=25mm]{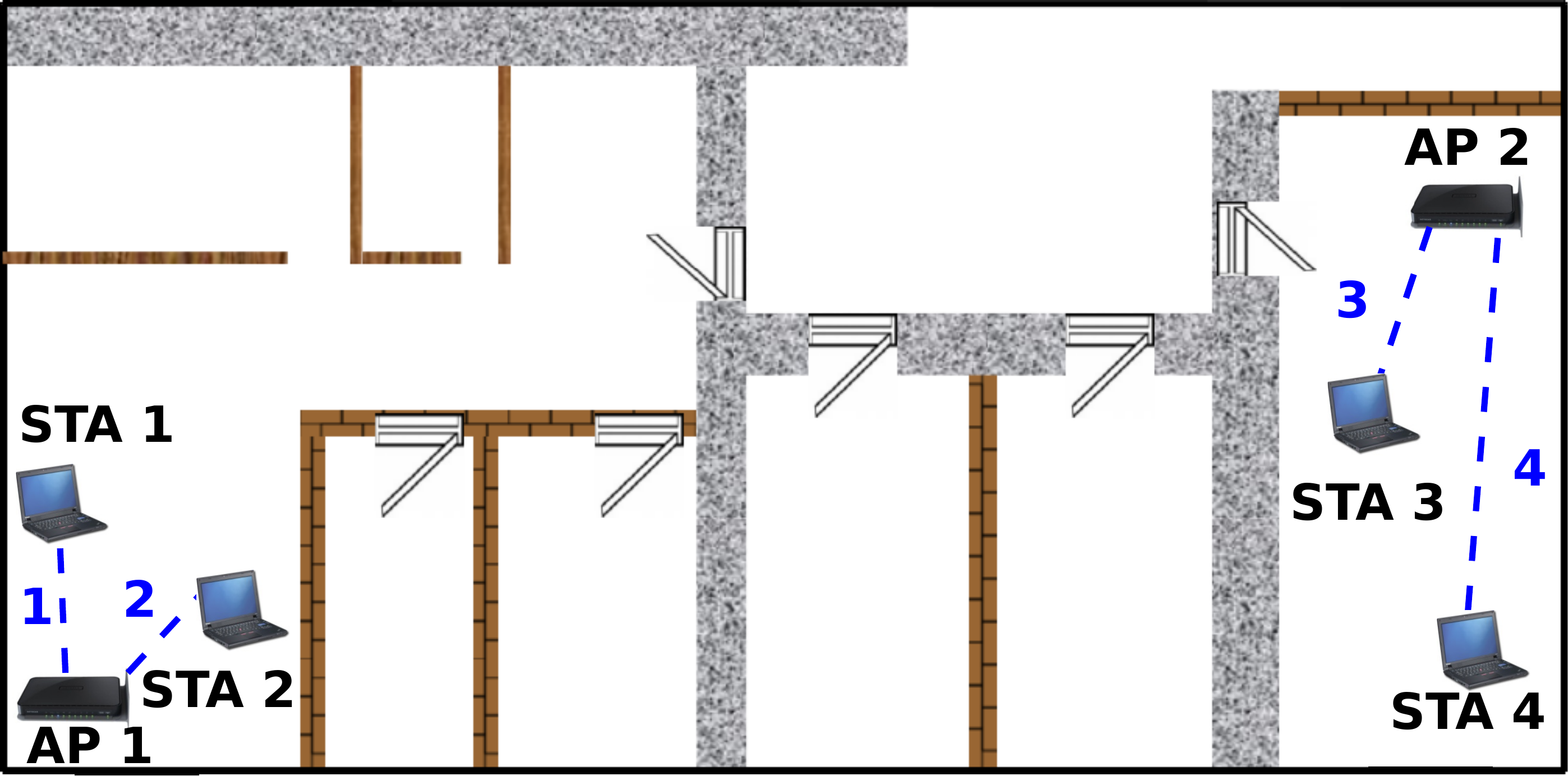}
		\caption{Physical position of clients. The dashed lines indicate the client-AP associations. \label{fig:s1_place}}
	\end{subfigure}
	\hspace{5mm}
	\begin{subfigure}[b]{0.45\textwidth}
		\centering
		\includegraphics[scale=0.4]{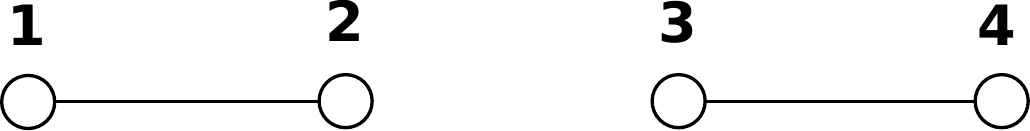}
		\caption{Dependence graph for the network in Fig.~\ref{fig:s1_place}. \label{fig:s1_graph}}
	\end{subfigure}
	\begin{subfigure}[b]{0.5\textwidth}
		\vspace{2mm}
		\centering
		\includegraphics[scale=0.4]{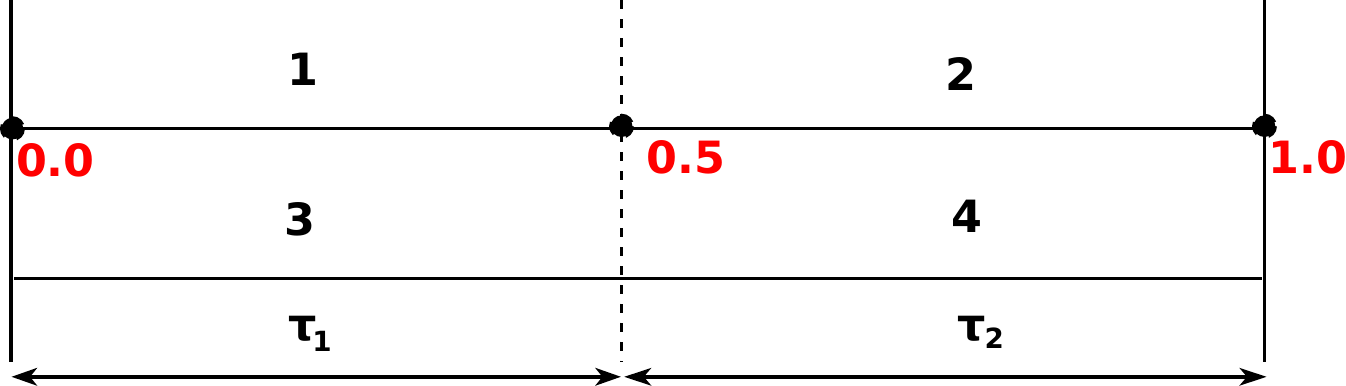}
		\caption{Time slice allocated, within a time-frame of duration $1000 \, ms$, to the clients in Fig.~\ref{fig:s1_place}. \label{fig:s1_slice}}
	\end{subfigure}
	\caption{Scenario 1 for 2 APs, 4 mobile clients; LAN-WLAN transfers.}
	\label{fig:s1}
\end{figure}

This experiment demonstrates the dependence inference and dynamic rate adaptation capabilities of ADWISER~v2. In this experiment, we consider four clients (STA~1, STA~2, STA~3 and STA~4) and two access points (AP1 and AP2). We follow a static association policy throughout this experiment, i.e., STA~1 and STA~2 are associated with AP1, whereas STA~3 and STA~4 are associated with AP2. All the clients are downloading large files from a LAN server, through their respective APs. Initially (see Fig.~\ref{fig:s1_place}), the clients are placed very close to the APs they are associated with. The dependence graph in this scenario as inferred by ADWISER~v2 is shown in Fig.~\ref{fig:s1_graph}. Since the APs do not interfere with each other, only clients associated with the same AP are dependent. The time slices obtained for this location is shown in Fig.~\ref{fig:s1_slice}.

\begin{figure}[!ht]
	\centering	
	\begin{subfigure}[b]{0.45\textwidth}
		\centering
		\includegraphics[height=25mm]{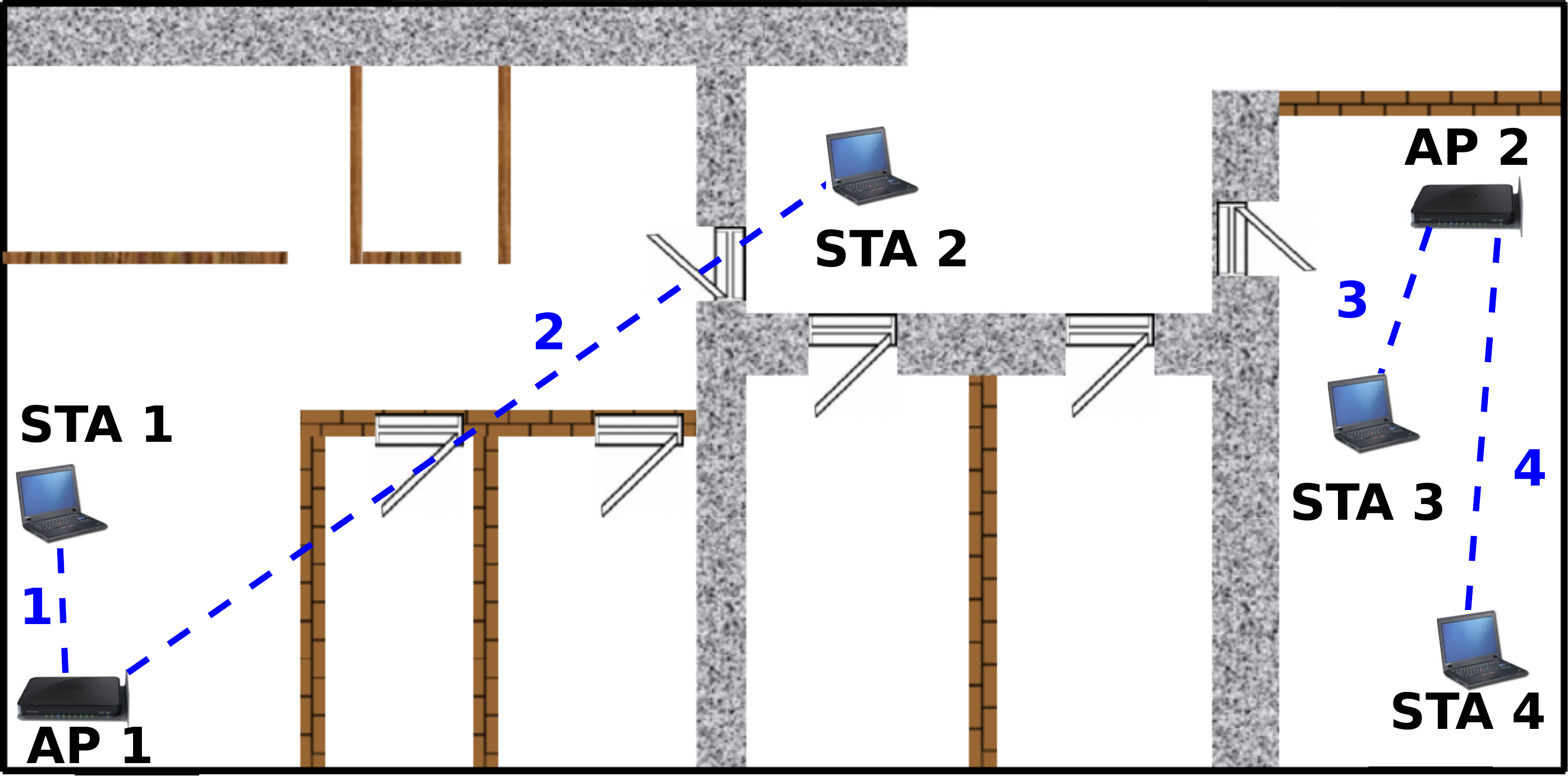}
		\caption{Physical position of clients. The dashed lines indicate the client-AP associations. \label{fig:s2_place}}
	\end{subfigure}
	\hspace{5mm}
	\begin{subfigure}[b]{0.45\textwidth}
		\centering
		\includegraphics[scale=0.4]{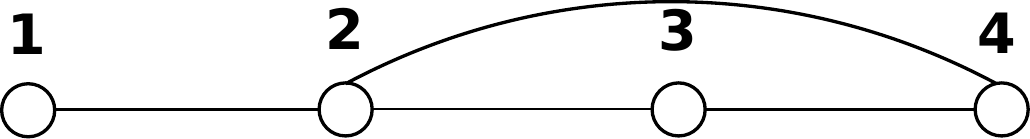}
		\caption{Dependence graph for the network in Fig.~\ref{fig:s2_place}. \label{fig:s2_graph}}
	\end{subfigure}
	\begin{subfigure}[b]{0.5\textwidth}
			\vspace{2mm}
		\centering
		\includegraphics[scale=0.4]{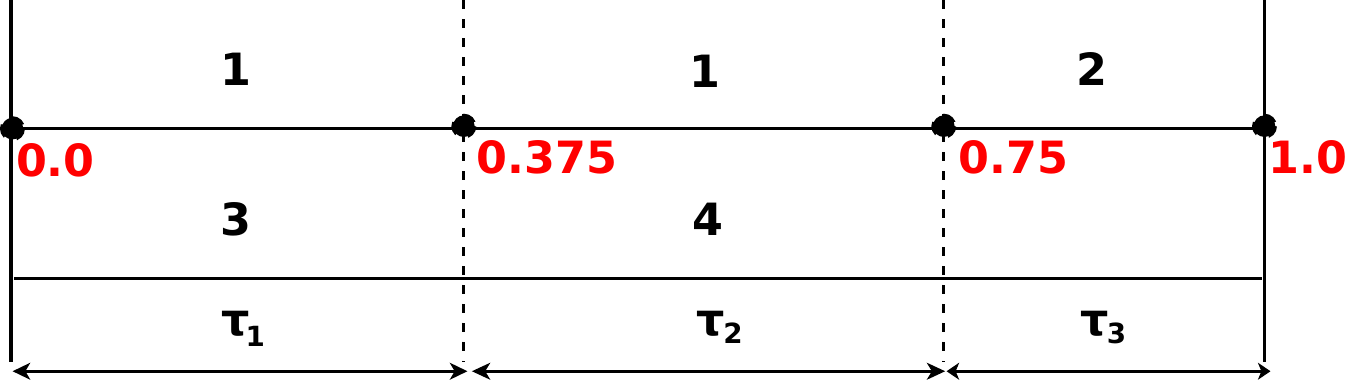}
		\caption{Time slice allocated, within a time-frame of duration $1000 \, ms$, to the clients in Fig.~\ref{fig:s2_place}. \label{fig:s2_slice}}
	\end{subfigure}
	\caption{Scenario 2 for 2 APs, 4 mobile clients; LAN-WLAN transfers afer moving STA~2. \label{fig:s2}}
\end{figure}
\begin{figure}[!ht]
	\centering	
	\begin{subfigure}[b]{0.45\textwidth}
		\centering
		\includegraphics[height=25mm]{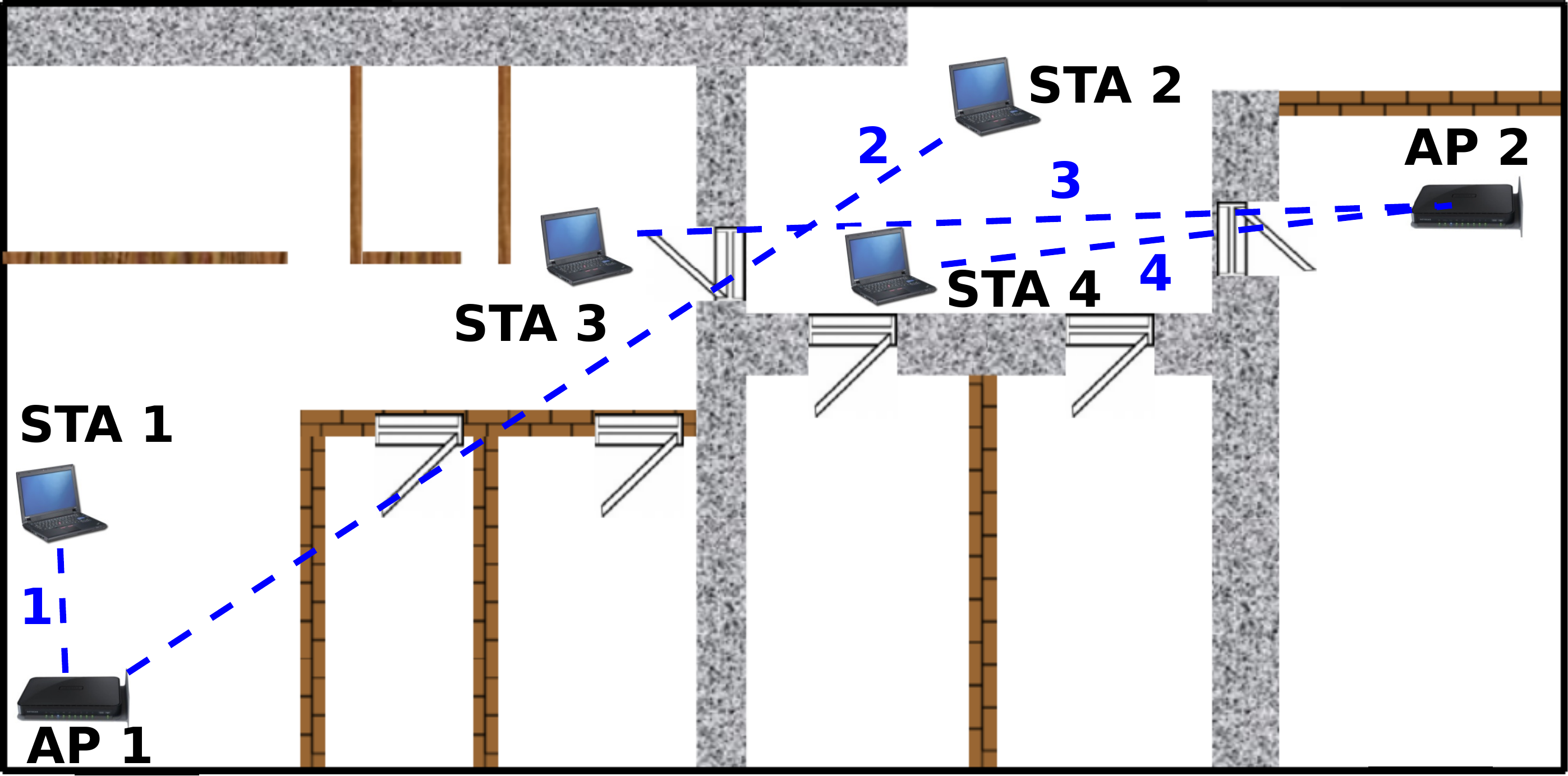}
		\caption{Physical position of clients. The dashed lines indicate the client-AP associations. \label{fig:s3_place}}
	\end{subfigure}
	\hspace{5mm}
	\begin{subfigure}[b]{0.45\textwidth}
		\centering
		\includegraphics[scale=0.4]{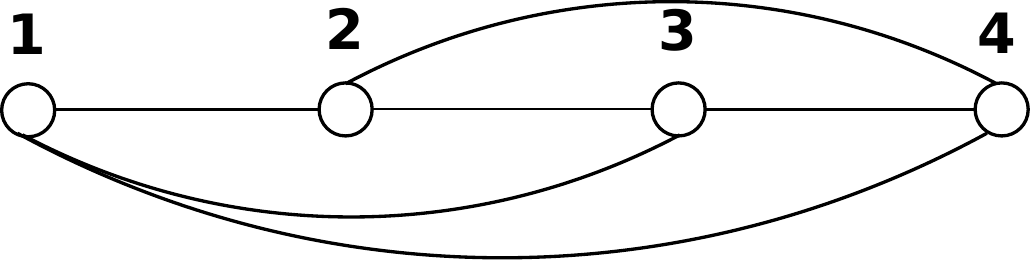}
		\caption{Dependence graph for the network in Fig.~\ref{fig:s3_place}. \label{fig:s3_graph}}
	\end{subfigure}

	\begin{subfigure}[b]{0.5\textwidth}
			\vspace{2mm}
		\centering
		\includegraphics[scale=0.4]{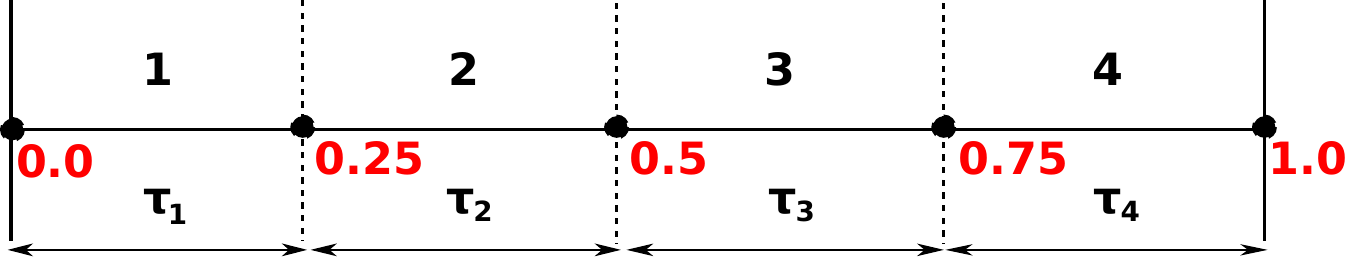}
		\caption{Time slice allocated, within a time-frame of duration $1000 \, ms$, to the clients in Fig.~\ref{fig:s3_place}. \label{fig:s3_slice}}
	\end{subfigure}
	\caption{Scenario 3 for 2 APs, 4 mobile clients; LAN-WLAN transfers afer moving STA~2. \label{fig:s3}}
\end{figure}

\begin{figure}[!ht]
		\centering
		\includegraphics[scale=0.33,angle=-90]{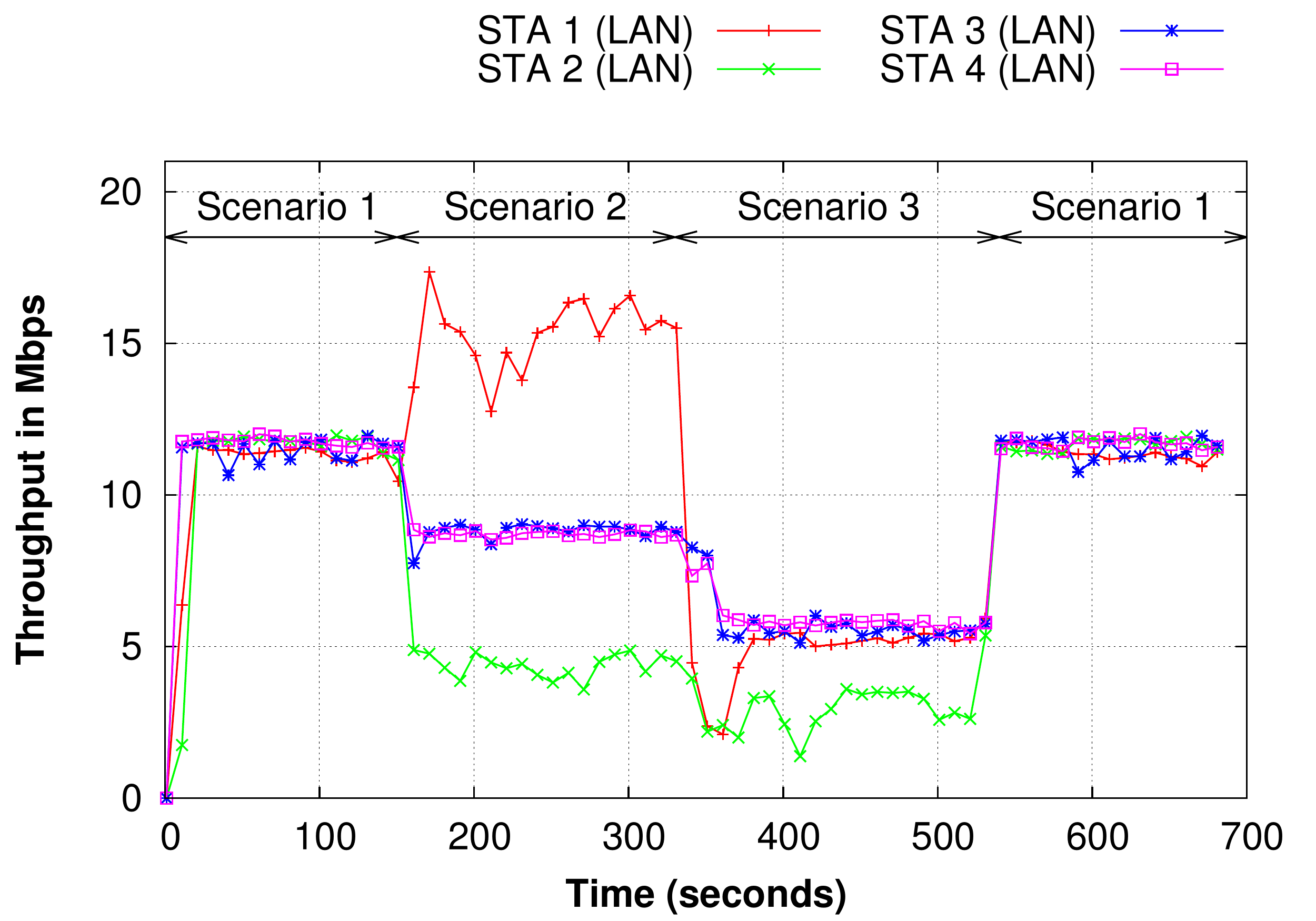}
		\caption{2 APs, 4 mobile clients; LAN-WLAN transfers (experimental setup in Fig.~\ref{fig:s1}, Fig.~\ref{fig:s2} and Fig.~\ref{fig:s3}): Throughput of the clients, when the TCP connections are managed by ADWISER~v2.}
		\label{fig:dld_ra_2ap_4sta}
\end{figure}

After about $150 \, seconds$, STA~2 is moved to the position shown in Fig.~\ref{fig:s2_place}. ADWISER~v2 automatically updates the link dependence graph to capture the new dependencies (see Fig.~\ref{fig:s2_graph}). ADWISER~v2 also adapts the virtual server service rate of the clients to reflect the physical rate of association of the clients. The time slices for this scenario are shown in 
Fig.~\ref{fig:s2_slice}. At about $350 \, seconds$, STA~3 and STA~4 are moved physically closer to AP1 (see Fig.~\ref{fig:s3_place}). Then, all the clients are declared to be dependent on one another by ADWISER~v2 (see Fig.~\ref{fig:s3_graph}). The corresponding time slices are shown in Fig.~\ref{fig:s3_slice}. Finally, after about $550 \, seconds$, all the clients are  brought back to their original position. Fig.~\ref{fig:dld_ra_2ap_4sta} shows the throughput of the clients obtained in the three scenarios. 

\begin{figure}[!ht]
 	\centering
 	\includegraphics[scale=0.35]{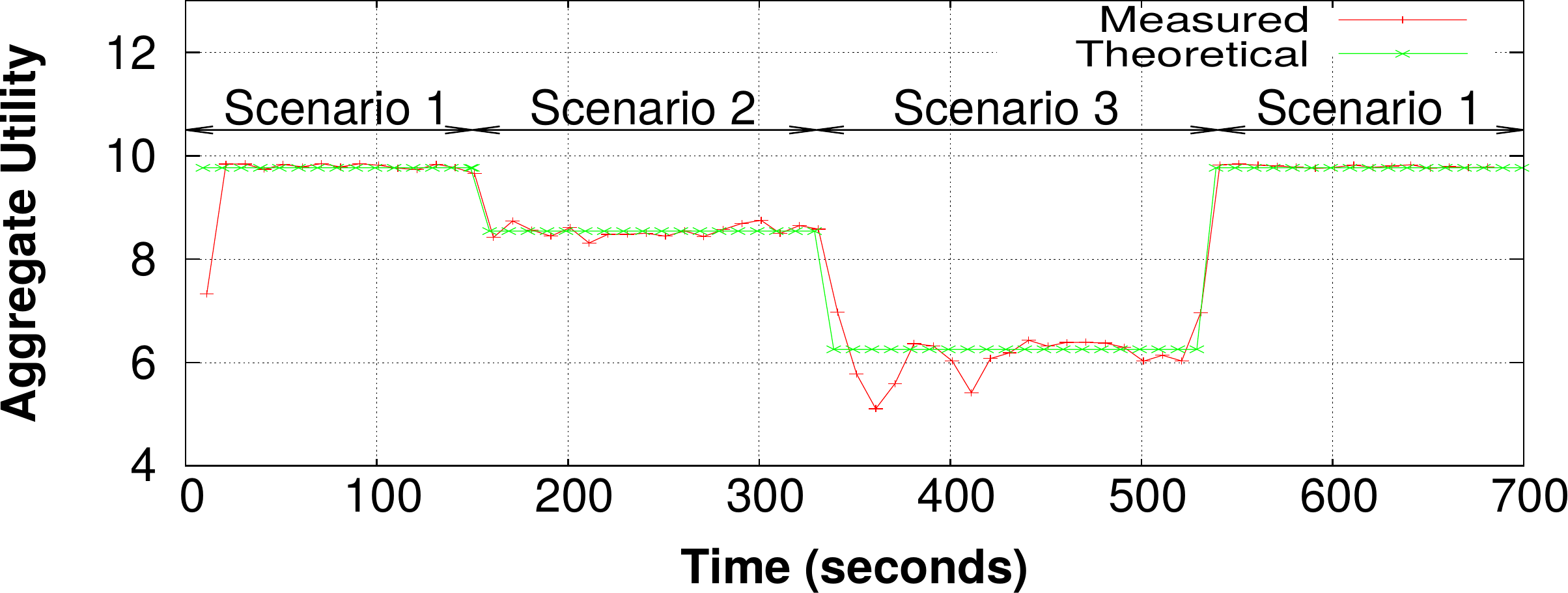}
 	\caption{2 APs, 4 mobile clients; LAN-WLAN transfers (experimental setup in Fig.~\ref{fig:s1}, Fig.~\ref{fig:s2} and Fig.~\ref{fig:s3}): Plot showing the theoretical 
 		and measured aggregate utility.}
 	\label{fig:dld_ra_2ap_4sta_utility}   	 
 \end{figure}

From Fig.~\ref{fig:dld_ra_2ap_4sta}, we can see that the throughputs of the clients are quite flat over time, and are proportional to the total WLAN time allocated to the clients.  We compute the theoretical aggregate utility (see Section \ref{sec:utility-optimization}) for each of the scenarios, and compare it with the measured aggregate utility in Fig.~\ref{fig:dld_ra_2ap_4sta_utility}. From Fig.~\ref{fig:dld_ra_2ap_4sta_utility}, we can see that the theoretical and measured aggregate utility are in close agreement with each other.  Thus demonstrating the ability of ADWISER~v2 to dynamically match the theoretical aggregate utility.

\section{Managing Short-lived and Interactive TCP traffic }
\label{sec:slit}

\begin{figure}[ht]
  	\centering
  	\includegraphics[height=25mm]{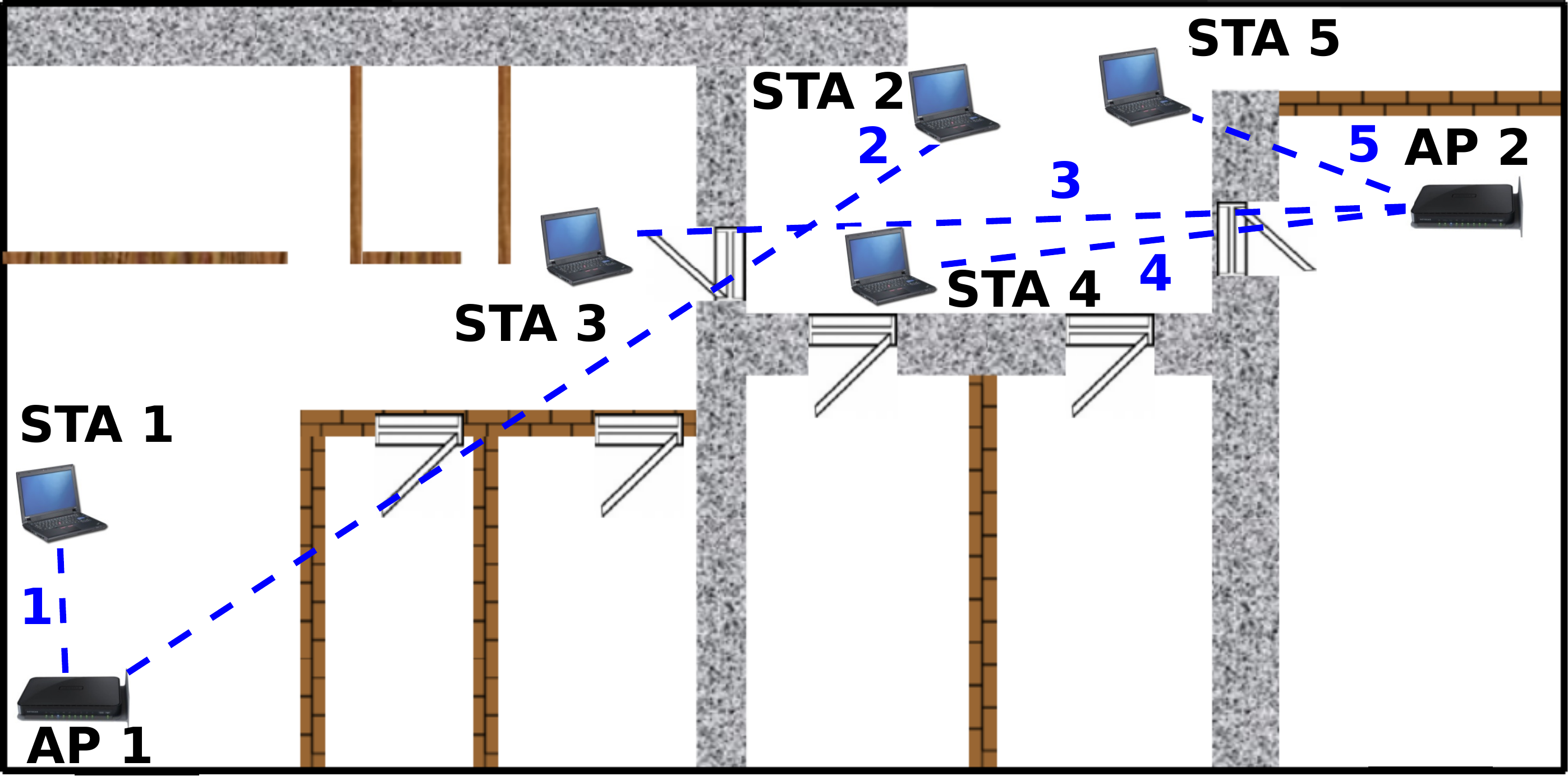}
  	\caption{2 APs, 5 clients; coexisting TCP transfers: Physical position of clients. The dashed lines indicate the client-AP associations.}
  	\label{fig:sl_place}
  	  \end{figure}

\begin{figure}[ht]
  	 	\centering
  	 	\includegraphics[scale=0.4]{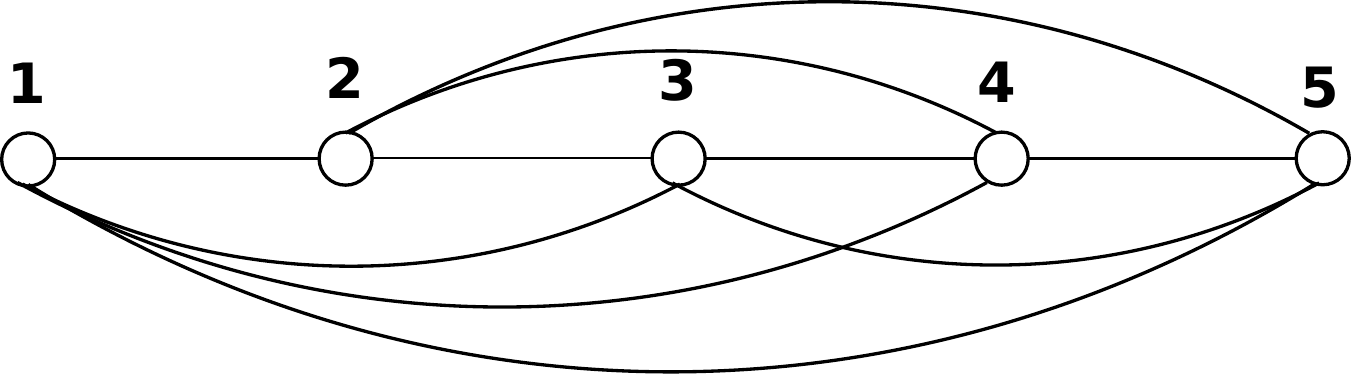}
  	 	\caption{2 APs, 5 clients; coexisting TCP transfers: Dependence graph for the network in Fig.~\ref{fig:sl_place}.}
  	 	\label{fig:sl_graph}
  \end{figure}

 In this section, experimentally, we study the performance of short-lived and interactive TCP traffic, in the presence of long-lived TCP transfers. The experimental setup for this section and the corresponding dependence graph are shown in Fig.~\ref{fig:sl_place} and Fig.~\ref{fig:sl_graph}, respectively.
 
  
 \subsection{Short-lived TCP transfers}  \label{sec:sl}
  Though long-lived TCP transfers constitute a large portion of the total traffic generated by the clients, the presence of coexisting short-lived TCP transfers and their impact on the throughput of long-lived TCP transfers need to be considered when developing and implementing any WLAN performance management solution. In this section, experimentally, we study the interactions between short-lived and long-lived TCP transfers on the WLAN. Further, we also demonstrate how a simple tweak to our time-slicing approach can help us accommodate and isolate short-lived and long-lived transfers.

\begin{figure}[ht]
  	\centering
  	\includegraphics[scale=0.3, angle=-90]{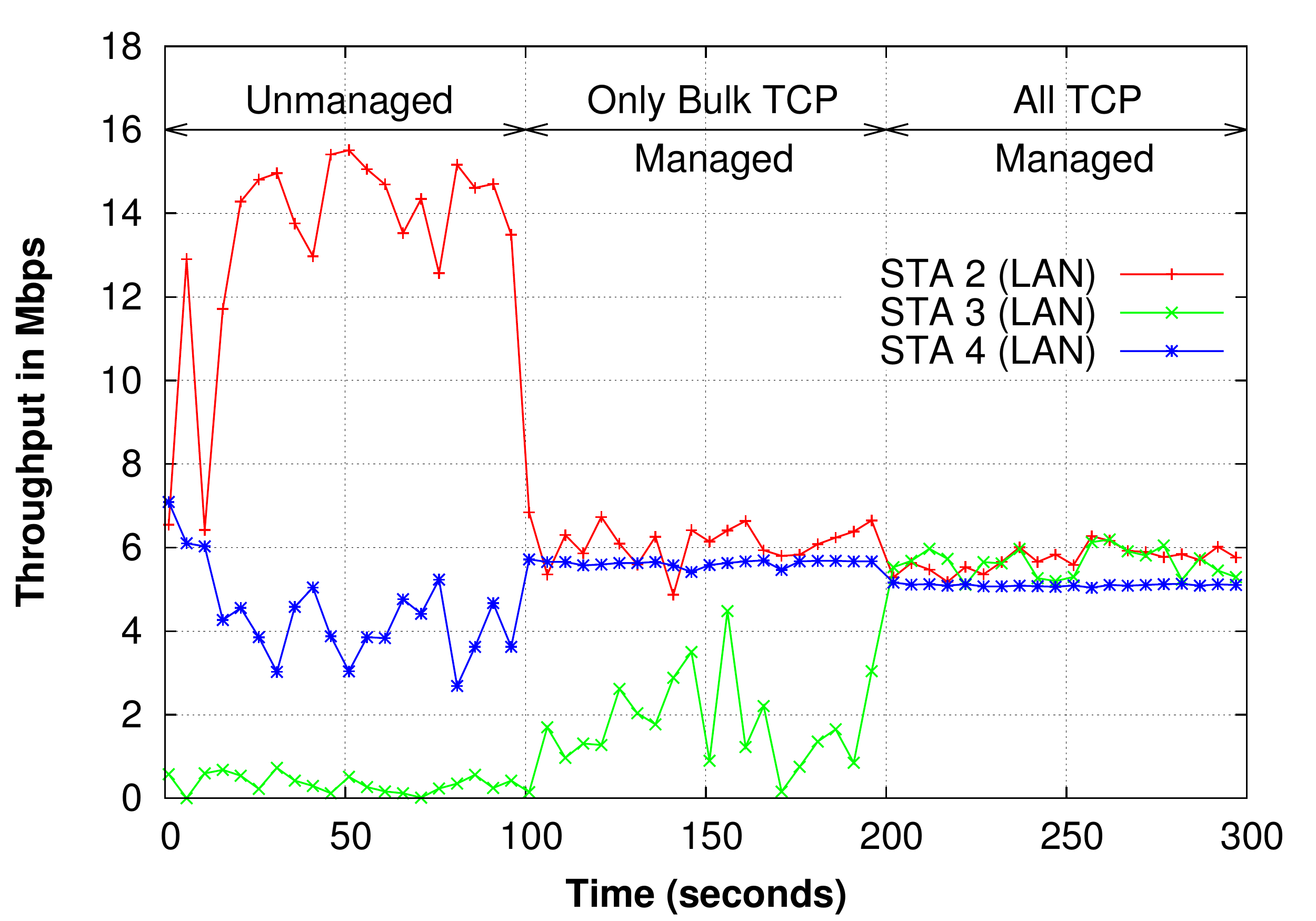}
  	\caption{2 APs, 5 clients; coexisting short-lived and long-lived TCP transfers (experimental setup in Fig.~\ref{fig:sl_place}): Throughput of clients doing long-lived download.}
  	\label{fig:s4_web_thruput}
  \end{figure}

\begin{figure}[ht]
  	\centering
  	\includegraphics[scale=0.35]{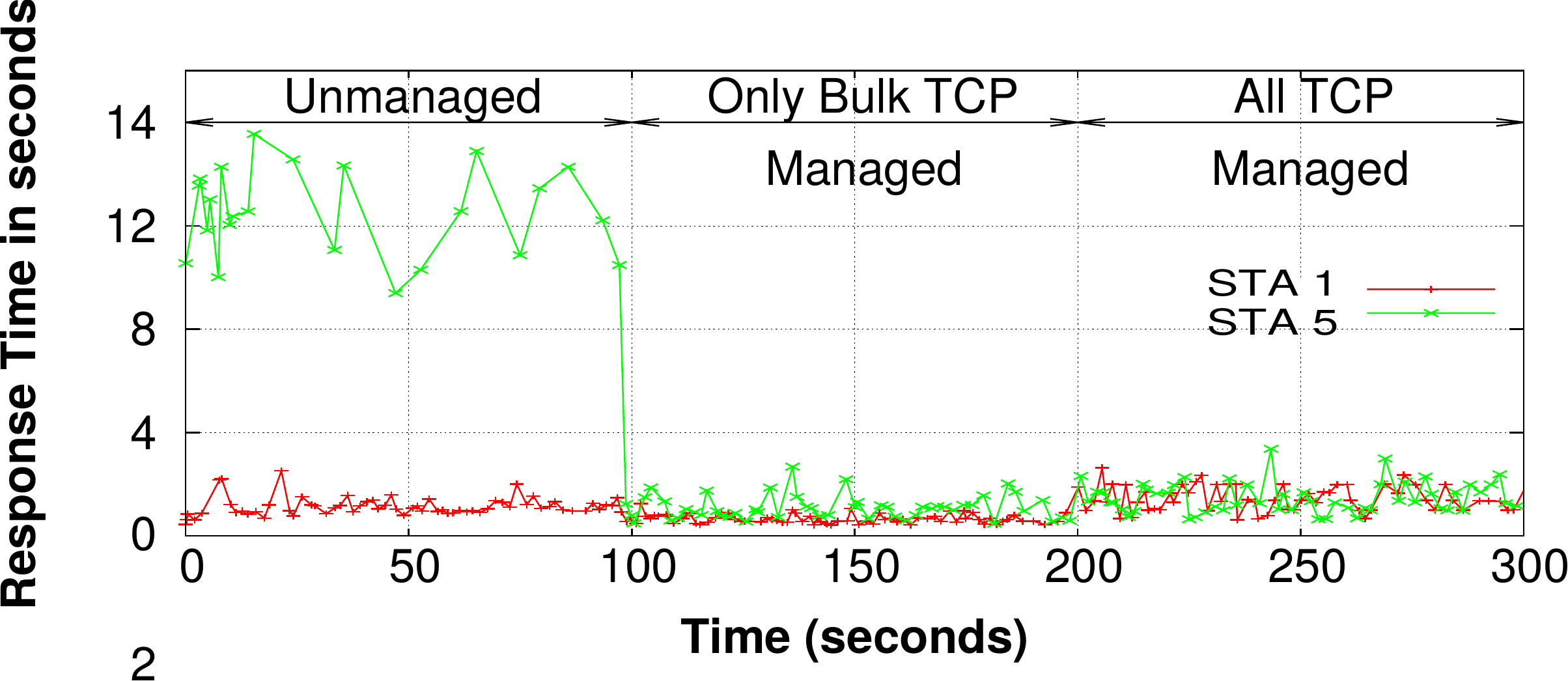}
  	\caption{2 APs, 5 clients; coexisting short-lived and long-lived LAN-WLAN transfers (experimental setup in Fig.~\ref{fig:sl_place}): Response times of the clients performing short-lived TCP transfer. }
  	\label{fig:s4_web_response_time}
\end{figure}

  Consider the scenario shown in Fig.~\ref{fig:sl_place}. To begin with, we perform an experiment in which STA~2, STA~3 and STA~4 are downloading large files from a LAN server, and STA~1 and STA~5 are performing short-lived file transfers (web like traffic) from a WAN server with RTPD of $150 \, ms$, through the proxy. From $0-100 \, seconds$, we run the network in ``unmanaged mode,'' i.e., without time-slicing. From Fig.~\ref{fig:s4_web_thruput}, we can see that in the \emph{unmanaged mode}, with the five clients contending simultaneously, STA~3 obtains a very low throughput (almost zero). Also, STA~2 and STA~4 obtain highly variable throughputs of about $15 \, Mbps$ and $4 \, Mbps$, respectively. Further, due to heavy contention from co-channel APs and clients, the \emph{unmanaged mode} also results in large response times for STA~5 (see Fig.~\ref{fig:s4_web_response_time}). 
  
  \begin{figure}[ht]
   	\centering
   	\includegraphics[scale=0.5]{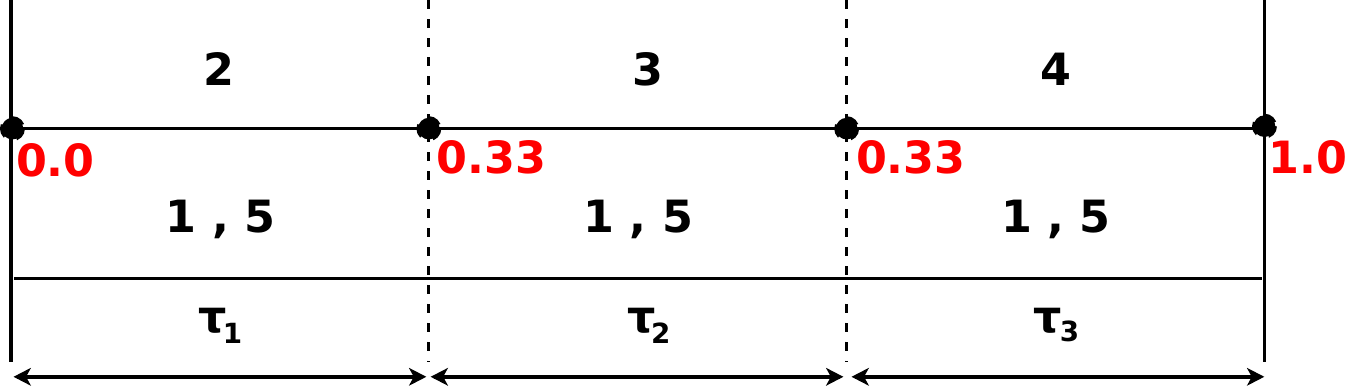}
   	\caption{2 APs, 5 clients; coexisting short-lived and long-lived TCP transfers (experimental setup in Fig.~\ref{fig:sl_place}): \textbf{Only long-lived/bulk TCP transfers are time sliced}; time slices from $100$ to $200 \, seconds$.}
   	\label{fig:s4_web_ll_ts}
   \end{figure}
   
   Next, for $100 - 200\, seconds$, we time slice only long-lived TCP transfers, i.e., (STA~2, STA~3 and STA~4). During this period, each of the long-lived TCP transfers is served in non-overlapping times slices of duration $333 \, ms$, and the clients doing short-lived file transfers are served in all the time slices (see Fig.~\ref{fig:s4_web_ll_ts}).  The coarse time-slicing of clients performing long-lived transfers shifts the queues from the APs to ADWISER~v2. This in turn reduces the contention faced by the clients that are served all the time (i.e., STA~1 and STA~5), thereby drastically improving the response time of the clients doing short-lived file transfers (see Fig.~\ref{fig:s4_web_response_time}). While each of STA~2 and STA~4 enjoy a fairly flat throughput of $6 \, Mbps$ in $100 - 200\, seconds$, STA~3 still obtains a low throughput of about $2 \, Mbps$. Since STA~3 is time-sliced, its queue at the AP is almost empty. Hence packets destined to STA~3 have to contend with those destined to STA~5 (since both clients belong to same AP). On top of that, STA~3 also faces significant interference due to transmissions from AP1. All these reasons result in STA~3 obtaining a low and highly variable throughput. 	
  
  \begin{figure}[ht]
  	\centering
  	\includegraphics[scale=0.5]{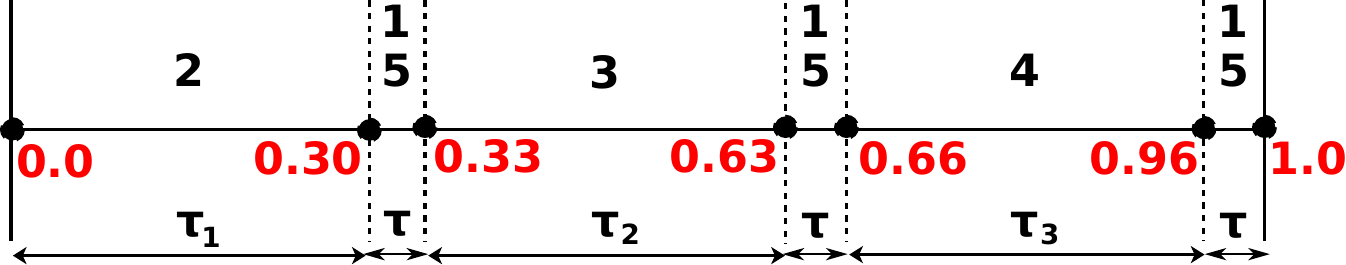}
  	\caption{2 APs, 5 clients; coexisting short-lived and long-lived TCP transfers (experimental setup in Fig.~\ref{fig:sl_place}): \textbf{All TCP transfers are time sliced}; time slices from $200$ to $300 \, seconds$.}
  	\label{fig:s4_web_all_ts}
  \end{figure}
  
From $200 \,seconds$ to $300\, seconds$, we time slice all the TCP transfers i.e., (transfers to all the five clients). To achieve this, in a time time-frame of length one second, we allocate a duration of $100 \, ms$ to the clients performing short-lived TCP transfers (STA~1 and STA~5), and divide the remaining $900 \, ms$ equally among the other clients (each will get $300 \, ms$). Essentially, 
this is equivalent to isolating the short-lived and long-lived file transfers. Further, we split the time slice allocated to STA~1 and STA~5 into three equal parts and space them equally within a time-frame (see Fig.~\ref{fig:s4_web_all_ts}). While the management of short-lived file transfers results in a slight increase in the response time of STA~1 and STA~5 (see Fig.~\ref{fig:s4_web_response_time}), it leads to reduced interference, thereby improving the throughput of STA~3 (see Fig.~\ref{fig:s4_web_thruput}). 

While we have demonstrated a coarse time-sliced scheme where  long-lived and short-lived file transfers can co-exist without affecting each other, the slice allocation scheme presented in this section need not be optimal. We plan to address the problem of finding the optimal time slice allocation scheme for coexisting long and short lived TCP transfer in our future work.

\subsection{Interactive TCP traffic}
\label{sec:it}

Another class of popular applications is that of interactive applications (e.g., VoIP, SSH). Such applications have low bandwidth and  low delay requirements. Since the bandwidth requirement of such applications is almost negligible, packets from such applications can be bypassed altogether (i.e., the packets are served immediately, irrespective of the time-sliced schedule). From the previous section, it is clear that unmanaged long-lived TCP transfers can drastically affect the response times of other TCP transfers. Therefore, it is crucial that we qualitatively evaluate the effect of coarse time-slicing on traffic with low delay requirement. The experimental setup for this section and the corresponding dependence graph are shown in Fig.~\ref{fig:sl_place} and Fig.~\ref{fig:sl_graph}, respectively.

\begin{figure}[ht]
	\centering
	\includegraphics[scale=0.35]{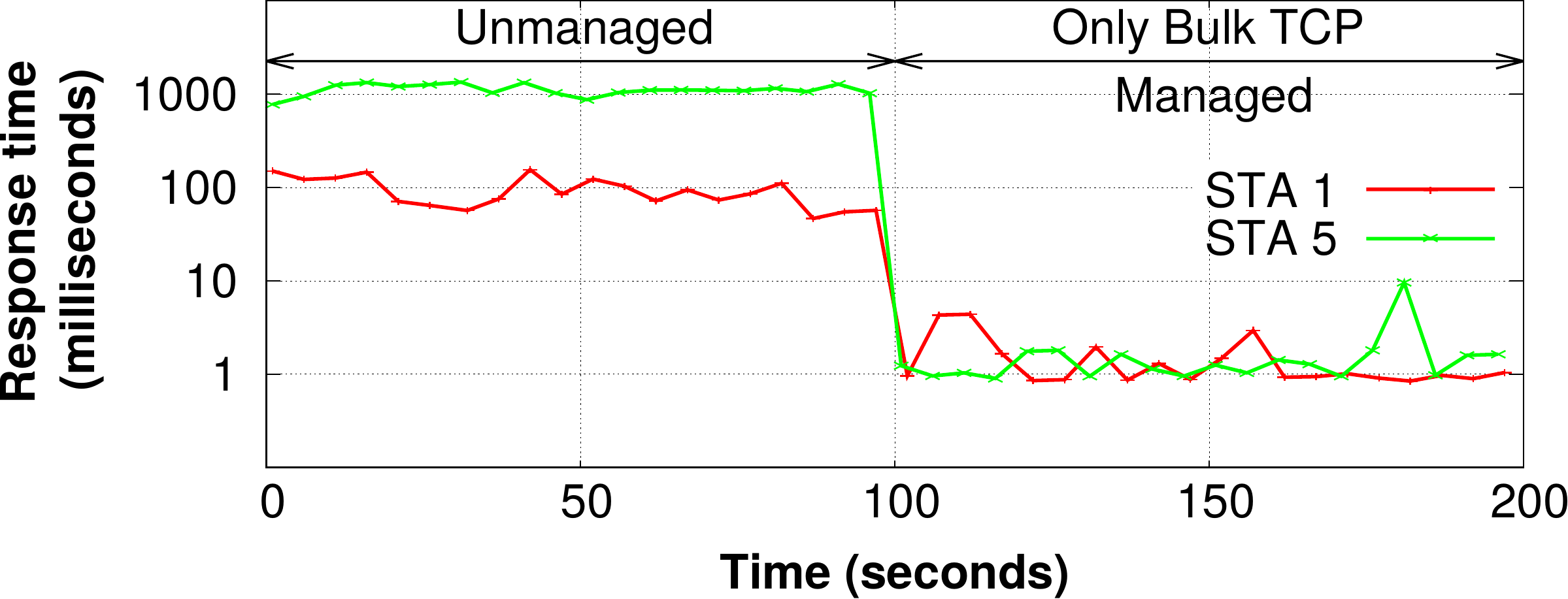}
	\caption{2 APs, 5 clients; coexisting interactive and long-lived traffic: Ping response times with and without ADWISER~v2.}
	\label{fig:s5_ping}
\end{figure}

In this experiment, as in the previous section, STA~2, STA~3 and STA~4 are downloading large files from a server on the LAN. To observe the effect of these long-lived TCP transfers on the delay of low bandwidth application, ping packets of size $64 \, Bytes$ are sent from the same LAN server to STA~1 and STA~5, at regular  intervals of $1 \, second$.  From $0-100 \, seconds$, the system is run in the \emph{unmanaged mode}. In this mode, the default IEEE~802.11 DCF behaviour dictates the performance of the network. Since long-lived file transfers are not throttled, their queues almost always reside in the APs. As a consequence of this, the ping packets destined to STA~1 face heavy contention from those destined to STA~2, and the ping packets sent to STA~5 face contention from STA~3 and STA~4, thereby resulting in large RTPD for the ping packets (see Fig.~\ref{fig:s5_ping}).

For $100-200 \, seconds$, we time slice all the long-lived TCP transfers (i.e., STA~2, STA~3 and STA~4). 
Since the bandwidth requirement for the ping packet is very small, they are served as soon as they arrive at ADWISER~v2. The time slices for this experiment is shown in Fig.~\ref{fig:s4_web_ll_ts}.  Since there are three clients doing long-lived downloads, the solution of the optimization problem discussed in Section \ref{sec:utility-optimization} requires us to allocate equal time slices to each of the clients (i.e., $333 \, ms$ each). The RTPD experienced by the ping packets when the long-lived TCP transfers are managed is presented in Fig.~\ref{fig:s5_ping}. From Fig.~\ref{fig:s5_ping} we can see that coarse time slicing of the long-lived TCP transfers drastically improves the delay experience of the ping packets from several hundreds of milliseconds to just a few a few milliseconds. 


	\section{An Experiment with IEEE~802.11n WLAN}
	\label{sec:11n}

	IEEE~802.11n based WLANs have become popular in recent years. The IEEE~802.11n standard extends the IEEE~802.11g WLAN standard by significantly increasing reach, reliability, and throughput. It offers a significant increase in the maximum physical rate of association from $54 \, Mbps$ to $600 \,Mbps$ with the use of four spatial streams at a channel width of $20 MHz$. IEEE 802.11n standardizes support for multiple-input multiple-output (MIMO) and frame aggregation, among other features. To demonstrate the versatility of ADWISER~v2, and to highlight the technology agnostic techniques adopted by ADWISER~v2, we perform an experiment on an IEEE~802.11n infrastructure network.

\begin{figure}[ht]
			\centering
			\includegraphics[scale=0.33, angle=-90]{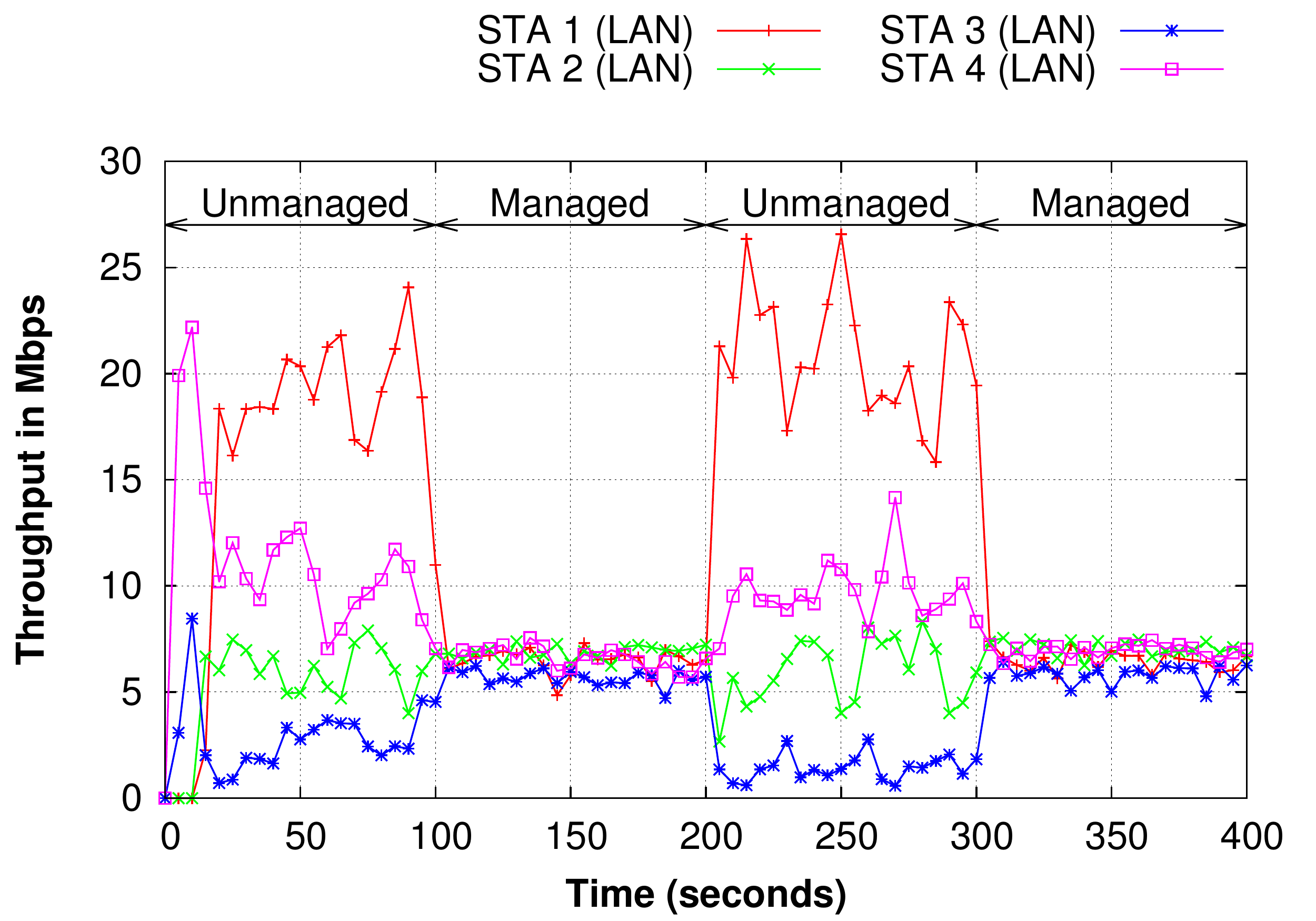}
			\caption{2 IEEE~802.11n APs, 4 clients; LAN Downloads (experimental setup in Fig.~\ref{fig:s2_place}): Throughput obtained of the client.}
			\label{fig:11n}
		\end{figure}

	The physical placement of the clients, in this experiment, is shown in Fig.~\ref{fig:s2_place}. The only difference is that, in this experiment, the APs and clients are configured to use \emph{IEEE~802.11n}. The dependence graph and time slices allocated to the clients are  shown in Fig.~\ref{fig:s2_graph} and Fig.~\ref{fig:s2_slice}, respectively. All the clients are downloading large files from a LAN server, through their respective associated APs. The throughputs obtained by the clients in the \emph{managed} and \emph{unmanaged} modes are presented in Fig.~\ref{fig:11n}. 
	
	From Fig~\ref{fig:11n}, we can see that, with time-slicing (i.e., in the \emph{managed mode}), all the clients enjoy equal throughputs of about $7 \, Mbps$ each. On the other hand, in the \emph{unmanaged mode}, the throughput of STA~1 shoots up to $25 \, Mbps$, and the throughput of STA~3 drops to about $2 \, Mbps$. The main reason for the reduction in the throughput of STA~3 is interference from uncontrolled transmissions destined to STA~1. Since IEEE~802.11n standard supports new features like MIMO and frame aggregation, IEEE~802.11n based WLANs may exhibit performance anomalies that we have not encountered so far. We plan to study such performance anomalies in our future work. 

\section{Conclusion}
\label{sec:conclusion}

In this paper, we have reported our control approach and experiments with ADWISER~v2 --- a WLAN QoS controller aimed at addressing performance issues for long-lived TCP transfers in WLANs with interfering co-channel APs.
ADWISER~v2 overlays coarse time-slicing on top of a cascaded fair queuing scheduler, to achieve considerably better
and controllable TCP performance without any modification to the firmware/hardware of the APs and clients. The time slices
have been obtained by solving a constrained utility maximization problem.
We have demonstrated the
efficacy of coarse time-slicing of TCP connections, combined with a TCP proxy,
in managing transfers. 
Our experimental results show that the contrast between the unmanaged
and managed modes of operation is remarkable: From highly variable individual
throughputs and poor aggregate rates, the system moves to a regime with stable throughputs and higher aggregate rates. ADWISER~v2 is able to ensure \emph{fair and efficient operation of multi-AP WLANs, without any change to hardware and firmware of the 
network devices.}

\section*{References}
\bibliography{adwiser_els_cn}
\end{document}